\documentclass[aps,superscriptaddress,showpacs,amsmath,amssymb,amsfonts,altaffillsymbol,notitlepage,pre]{revtex4-2}
\usepackage[dvips]{graphicx}
\usepackage{color}
\usepackage{soul}
\usepackage{epsfig}
\usepackage{bm}
\usepackage[normalem]{ulem}
\usepackage[utf8]{inputenc}
\usepackage[T1]{fontenc}
\usepackage{mathptmx}
\usepackage{etoolbox}
\usepackage{natbib}

\usepackage{placeins}
\usepackage[caption=false,font=normalsize,labelfont=bf]{subfig}

\begin{document}


\title{Collective spatial reorganization from arrest to peeling and migration through density-dependent mobility in internal-state coordinates}

\author{Yagyik Goswami}

\affiliation{$^{*}$Laboratory of Multiscale Bioimaging, Paul Scherrer Institute, Forschungstrasse 111, Villigen, Aargau, 5232, Switzerland}



\email{yagyik.goswami@psi.ch}


\begin{abstract}
Numerous problems in development, regeneration, and disease involve simultaneous evolution of both spatial organization and the internal state of the constituents in addition to local interactions and crowding. This motivates us to study a minimal model for interacting populations evolving in coupled spatial and internal-state coordinates. We focus on a specific transition of particular biological interest: the reorganization of dense collectives from compact or arrested states toward boundary-led peeling and migration. In our formulation, each particle carries a spatial position and a scalar internal state, and interacts through finite-range forces. Mobilities are defined on both spatial and internal-state coordinates with a density dependence, and are asymmetrically cross-coupled. We derive update equations for stochastic dynamics in the overdamped limit and perform numerical simulations. We find that mobility in internal-state coordinates alone provides an independent control axis for large-scale spatial reorganization. In particular, increasing the baseline internal-state diffusivity and tuning its density dependence drives a transition from arrested aggregates to a peeling-like regime with broad spatial excursions, strong outward radial bias, and edge-localized activity, while the baseline positional diffusivity is held fixed. The transition is accompanied by correlated broadening of spatial and internal-state displacements, systematic reorganization of radial density and density-curvature profiles, and a pronounced dependence on system size, consistent with the idea that growing aggregates can cross into a boundary-dominated migratory state. These results establish the utility of our approach and motivate a broader framework aimed at modeling state change, spatial redistribution, and neighborhood structure within a common formalism.
\end{abstract}

\maketitle

\section{Introduction}
\label{sec:introduction}

Spatial reorganization coupled to functional change is a recurring feature of living systems, from
development and tissue renewal to disease progression, engineered morphogenesis and metabolic specialization in microorganisms
\cite{oates2009quantitative,negrete2021towards,liu2024morphogenesis,seferbekova2023spatial,ackermann2023spatial}. In
many such settings, the central question is not only where cells are, but how state progression,
local neighbourhoods, and tissue architecture reshape one another over time. Recent advances in
spatial transcriptomics, multiplex imaging, lineage tracing, and image-based phenotyping now make it possible to observe these processes in increasingly joint form, measuring molecular state, local context, and spatial organization together rather than separately \cite{iber2016image,seferbekova2023spatial,bressan2023dawn,velten2023principles,rafelski2024establishing}.
At the same time, transport- and trajectory-based methods have made it much easier to represent
progression through high-dimensional state spaces
\cite{zhou2021dissecting,lange2022cellrank,bunne2024optimal}. This motivates a modelling framework in which physical position and internal state evolve as coupled
coordinates of a common dynamical system rather than as distinct post hoc descriptions.

A particularly important biological motif is the emergence of bulk--boundary differences. In
developing and self-organizing tissues, local signalling, crowding, curvature, and contact structure often differ sharply between dense interior regions and moving or reorganizing boundaries\cite{shahbazi2019self,uccar2021theory,seferbekova2023spatial}. Such differences suggest that transport coefficients could be treated as dependent on local density and that changes in internal-state mobility may feed back asymmetrically onto spatial motion. This is especially relevant for transitions between compact or arrested collectives and boundary-led reorganization, including peeling, escape, or migration-like behavior in stem-cell and organoid contexts \cite{warmflash2014method,nousi2021single}. 

A natural way to represent such coupled reorganization is to treat internal state as a continuous coordinate and to model transitions between states as stochastic dynamics on an effective free-energy-like landscape~\cite{zhou2021dissecting,teschendorff2021statistical,goswami2023kinetic,zhu2025quantifying}. In this view, spatial position and internal state are not separate descriptions but jointly evolving coordinates of the same interacting system. At the same time, biological organization is generically shaped by nonequilibrium effects~\cite{julicher2018hydrodynamic,shaebani2020computational,goswami2025yielding,sharma2025activity}, so the coupling between these coordinates need not be reciprocal or derivable from a purely passive gradient flow. This motivates studying minimal models in which density dependence, crowding, and asymmetric couplings can convert changes in internal-state mobility into qualitative changes in spatial organization.

Here we examine a minimal version of that idea. We study a particle model in coupled spatial and internal-state coordinates with density-dependent mobilities and an asymmetric cross-coupling
between the two coordinate sets. Our central question is whether mobility in internal-state coordinates alone can induce a transition in collective spatial organization while the baseline positional
diffusivity is held fixed. We show that indeed, varying the internal-state mobility drives a transition from compact or arrested aggregates to peeling-like outward reorganization and migration. This provides a useful starting point for studying a broad class of phenomena through dynamics in coupled spatial and internal-state coordinates, and a framework to build upon to study richer modes of collective organization such as alignment, memory, mechanics and data-driven inference. 

\section{Model}
\label{sec:model}

We consider an interacting population evolving in coupled spatial and internal-state coordinates, with physical position $\mathbf r_i\in\mathbb R^2$ and internal coordinate $ \mathbf {\tau_i}\in\mathbb R$ for each particle $i=1,\dots,N$. We begin by briefly outlining the general formulation, summarized by Eq.~\ref{eq:H_general_main}, and then specialize to the concrete model studied here: a two-dimensional overdamped $N$ particle system with finite-range interactions in real space, a continuous scalar internal state, density-dependent mobilities, and an explicitly asymmetric cross-coupling between spatial motion and local density curvature. This framework naturally connects to a continuity equation in the combined space of position and internal state; its relation to the microscopic equations of motion is derived in
Appendix~\ref{app:continuity_derivation}.
In the reciprocal passive limit, the stochastic forcing is constrained by the fluctuation-dissipation relation: the same transport coefficients that determine dissipative relaxation also determine the covariance of the noise, which is required for detailed balance and for sampling the correct stationary distribution of the passive system \cite{van1992stochastic}. In the present setting this generalizes to coupled spatial and internal-state coordinates, for which the stochastic increments are governed by a local covariance matrix built from the corresponding mobility matrix rather than by independent scalar noise strengths. We therefore treat the construction and sampling of this covariance carefully, and defer the full discussion to Appendix~\ref{app:noise_covariance}. Specific departures from the reciprocal passive case are then introduced in order to probe the implications of nonequilibrium dynamics; the resulting hierarchy of nonequilibrium structures in the model family is summarized in
Appendix~\ref{app:nonequilibrium_levels}.

\subsection{Hamiltonian}
\label{subsec:hamiltonian}

\noindent At the broadest level, the shared-coordinate framework begins from an effective Hamiltonian
\begin{equation}
\mathcal H
=
\sum_{i<j}
U\!\left(\mathbf r_i-\mathbf r_j,\tau_i,\tau_j\right)
+
\sum_i V_{\mathrm{intr}}(\tau_i),
\label{eq:H_general_main}
\end{equation}
where $U$ encodes pairwise interactions in the joint space of position and type, and
$V_{\mathrm{intr}}$ is a one-body potential in type space. In the present work, we specifically consider the following pair interaction:
\begin{equation}
V_{ij}(r,\Delta\tau)
=
\Phi_{\mathrm{rep}}(r)
+
J(\Delta\tau)\,\Phi_{\mathrm{shell}}(r)
+
C(\Delta\tau)\,\mathbf 1_{r<a}.
\label{eq:V_pair_main}
\end{equation}
The short-range repulsive core is Hertz-like with a size-parameter $a$,
\begin{equation}
\Phi_{\mathrm{rep}}(r)
=
\epsilon_{\mathrm{rep}}
\left(\frac{a-r}{a}\right)^{\alpha_{\mathrm{rep}}}
\Theta(a-r),
\label{eq:phi_rep_main}
\end{equation}
while the finite-range shell interaction for $a < r \leq r_c$ is
\begin{equation}
\Phi_{\mathrm{shell}}(r)
=
-\epsilon_{\mathrm{att}}
\Bigl[1-q\!\bigl(s(u)\bigr)\Bigr]
\Theta(r-a)\Theta(r_c-r),
\qquad
u=\frac{r-a}{r_c-a},
\qquad
s(u)=3u^2-2u^3,
\label{eq:phi_shell_main}
\end{equation}
with
\begin{equation}
q(z)
=
\frac{\tanh\!\bigl(\kappa(z-\tfrac12)\bigr)+\tanh(\kappa/2)}
{2\tanh(\kappa/2)}.
\label{eq:q_main}
\end{equation}
This construction gives a smooth shell with $\Phi_{\mathrm{shell}}(a)=-\epsilon_{\mathrm{att}}$,
$\Phi_{\mathrm{shell}}(r_c)=0$, and vanishing derivatives at both endpoints (see Fig.~\ref{fig:J0J2_overview}.
The shell amplitude is modulated by internal-state difference through
\begin{equation}
J(\Delta\tau)
=
J_0
+
J_2
\left(1-\frac{\Delta\tau^2}{\sigma_\tau^2}\right)
\Theta(\sigma_\tau-|\Delta\tau|),
\label{eq:J_main}
\end{equation}
where $J_0$ sets the baseline shell coupling, $J_2$ controls how similarity or contrast in
$\tau$ modifies the interaction, and $\sigma_\tau$ sets a scale for similarity/dissimilarity in internal state between individual cells. The contact offset
\begin{equation}
C(\Delta\tau)=J(\Delta\tau)\,\Phi_{\mathrm{shell}}(a)
\label{eq:C_offset_main}
\end{equation}
is included only for continuity of the potential at $r=a$ and does not affect positional forces. We show the potential schematically in Fig.~\ref{fig:J0J2_overview}.
The one-body type potential is taken to be a quartic double well,
\begin{equation}
V_{\mathrm{intr}}(\tau)
=
\frac{\lambda}{4}\tau^2(\tau-1)^2.
\label{eq:V_intr_main}
\end{equation}
At the continuum level, one can derive a free-energy functional on the joint density $\rho(\mathbf r,\tau,t)$ from Eq.~\eqref{eq:H_general_main}, through which one obtains the corresponding chemical potential and conservation law. The derivation for our specific case of interest, Eq.~\ref{eq:V_pair_main} is given in Appendix~\ref{app:continuity_derivation}. In the exact model studied here, the internal-state-specific interaction kernel is the scalar-$\tau$, giving us Eqs.~\eqref{eq:V_pair_main}--\eqref{eq:V_intr_main}.

\subsection{Equations of motion}
\label{subsec:dynamics}

\noindent The microscopic conservative forces are
\begin{equation}
\mathbf F^{(r)}_i
=
-\frac{\partial \mathcal H}{\partial \mathbf r_i},
\qquad
F^{(\tau)}_i
=
-\frac{\partial \mathcal H}{\partial \tau_i}.
\label{eq:forces_main}
\end{equation}
For the implemented pair potential these become
\begin{equation}
\mathbf F^{(r)}_i
=
\sum_{j\neq i}
\left[
\frac{d\Phi_{\mathrm{rep}}}{dr}(r_{ij})
+
J(\Delta\tau_{ij})
\frac{d\Phi_{\mathrm{shell}}}{dr}(r_{ij})
\right]
\frac{\mathbf r_{ij}}{r_{ij}},
\label{eq:F_r_main}
\end{equation}
and
\begin{equation}
F^{(\tau)}_i
=
-
\sum_{j\neq i}
\frac{dJ}{d(\Delta\tau)}(\Delta\tau_{ij})\,
\Phi_{\mathrm{shell}}(r_{ij})
-
\frac{dV_{\mathrm{intr}}}{d\tau_i},
\qquad
-\frac{dV_{\mathrm{intr}}}{d\tau}
=
-\lambda\,\tau(\tau-1)(2\tau-1).
\label{eq:F_tau_main}
\end{equation}
The associated continuum description is a continuity equation in the joint space of
$(\mathbf r,\tau)$,
\begin{equation}
\partial_t \rho(\mathbf r,\tau,t)
=
-\nabla_{\mathbf r}\!\cdot \mathbf J_r
-\partial_\tau J_\tau ,
\label{eq:continuity_main}
\end{equation}
with constitutive currents of Onsager form,
\begin{equation}
\begin{pmatrix}
\mathbf J_r \\[3pt]
J_\tau
\end{pmatrix}
=
-\rho
\begin{pmatrix}
\mathbf M_{rr} & \mathbf M_{r\tau} \\[3pt]
\mathbf M_{\tau r} & M_{\tau\tau}
\end{pmatrix}
\begin{pmatrix}
\nabla_{\mathbf r}\mu \\[3pt]
\partial_\tau \mu
\end{pmatrix}.
\label{eq:continuity_onsager_main}
\end{equation}

\noindent Defining the conservative drifts at particle level with density dependent mobilities (see Appendix~\ref{app:continuity_derivation} for the connection to continuum level description),
\begin{equation}
\mathbf v^{\mathrm{cons}}_{r,i}=M_r(\rho_i)\,\mathbf F^{(r)}_i,
\qquad
v^{\mathrm{cons}}_{\tau,i}=M_\tau(\rho_i)\,F^{(\tau)}_i,
\label{eq:cons_drifts_main}
\end{equation}
the stochastic equations of motion are,
\begin{align}
\dot{\mathbf r}_i
&=
\mathbf v^{\mathrm{cons}}_{r,i}
+
g_\chi\,M_{\tau r}(\rho_i)\,k_B T\,\widehat{\nabla^2\rho}_i\,
\widehat{\mathbf v}^{\mathrm{cons}}_{r,i}
+
\boldsymbol\eta_i(t), \nonumber \\
\dot{\tau}_i
&=
v^{\mathrm{cons}}_{\tau,i}
+
g_\chi\,M_{r\tau}(\rho_i)\,k_B T\,\widehat{\nabla^2\rho}_i
+
\zeta_i(t),
\label{eq:r_tau_dot_gen}
\end{align}
where
\begin{equation}
\widehat{\mathbf v}^{\mathrm{cons}}_{r,i}
=
\frac{\mathbf v^{\mathrm{cons}}_{r,i}}
{|\mathbf v^{\mathrm{cons}}_{r,i}|}
\qquad \text{for } |\mathbf v^{\mathrm{cons}}_{r,i}|>0.
\label{eq:vhat_main}
\end{equation}
In Eq.~\ref{eq:r_tau_dot_gen}, we have introduced a scaling parameter $g_{\chi}$ that modulates the cross terms of the drift, $M_{r\tau}$ and $M_{\tau r}$, in order to independently tune the strength of their input on the drift. The matrix components themselves are constrained by the validity of the noise covariance matrix in the passive baseline case, which is discussed in Appendix~\ref{app:noise_covariance}.

Additionally, and of key importance, introducing density-dependent mobilities necessitates a local density field for each particle. Further, this also necessitates a measure of local density curvature that quantitatively distinguishes bulk-like from boundary-like environments (see Appendix~\ref{app:local_density_curvature} for a discussion on how coarse-graining produces the Laplacian of the density as a proxy for curvature in Eq.~\ref{eq:r_tau_dot_gen} and Eq.~\ref{eq:r_tau_dot_main} below).
Thus, the cross-coupling acting in the positional equation is not the gradient of a type field, but a density-curvature signal projected along the instantaneous conservative positional drift. In the main simulations reported here, we consider the specific case of $M_{r\tau}=0$
so that the deterministic cross-coupling acts only on the positional components.  

\noindent The stochastic equations of motion become
\begin{align}
\dot{\mathbf r}_i
&=
\mathbf v^{\mathrm{cons}}_{r,i}
+
g_\chi\,M_{\tau r}(\rho_i)\,k_B T\,\widehat{\nabla^2\rho}_i\,
\widehat{\mathbf v}^{\mathrm{cons}}_{r,i}
+
\boldsymbol\eta_i(t), \nonumber \\
\dot{\tau}_i
&=
v^{\mathrm{cons}}_{\tau,i}
+
\zeta_i(t).
\label{eq:r_tau_dot_main}
\end{align}
Thus, for the specific case we study, internal-state dynamics evolves through its
conservative drift and stochastic forcing, while the positional dynamics receives in addition an
asymmetric density-curvature-driven cross drift projected along the instantaneous direction of the conservative forcing on position components. The positional reorganization is driven by a local bulk--rim contrast encoded in the density field. Alternative cross-coupling choices and their implications are discussed in Appendix~\ref{app:nonequilibrium_levels}.

The stochastic terms $\boldsymbol\eta_i(t)$ and $\zeta_i(t)$ denote zero-mean Gaussian forcing in the spatial and internal-state components, respectively. Their covariance is constructed from the same local mobility structure in Eq.~\ref{eq:continuity_onsager_main} that governs dissipative transport and is thus not chosen independently. In the passive limit, the fluctuation-dissipation relation holds, resulting in detailed balance being satisfied and ensuring that the correct stationary distribution is sampled. In the present coupled setting this leads to a local covariance matrix in the combined spatial and internal-state coordinates. We defer details of the explicit construction and numerical sampling of this covariance to Appendix~\ref{app:noise_covariance}.

\subsection{Mobility and density dependence}
\label{subsec:mobility_density}

\noindent To complete our description of the dynamics, we define for each particle a local density and a corresponding local density-curvature field from a smooth finite-range neighborhood average. In practice, the local density increases with the weighted number of nearby neighbors, while the density-curvature proxy measures whether a particle sits in a locally bulk-like or boundary-like environment. The explicit kernel construction is given in Appendix~\ref{app:local_density_curvature}.

These local fields define density-dependent transport, with our choice motivated by the
general observation that cells and other interacting constituents typically rearrange more slowly in crowded neighborhoods than at freer boundaries or interfaces~\cite{hallatschek2023proliferating,alert2020physical}. Accordingly, we let the positional and internal-state frictions increase with local density according to
\begin{equation}
\gamma_r(\rho)
=
\gamma_r^0
\left[
1+\left(\frac{\rho}{\rho_c}\right)^{\beta_r}
\right],
\qquad
\gamma_\tau(\rho)
=
\gamma_\tau^0
\left[
1+\left(\frac{\rho}{\rho_c}\right)^{\beta_\tau}
\right].
\label{eq:gamma_main}
\end{equation}
The corresponding diffusivities are
\begin{equation}
D_r(\rho)=\frac{k_B T}{\gamma_r(\rho)},
\qquad
D_\tau(\rho)=\frac{k_B T}{\gamma_\tau(\rho)},
\label{eq:diffusivities_main}
\end{equation}
and the overdamped mobilities are
\begin{equation}
M_r(\rho)=\frac{D_r(\rho)}{k_B T},
\qquad
M_\tau(\rho)=\frac{D_\tau(\rho)}{k_B T}.
\label{eq:diagonal_mobilities_main}
\end{equation}
Finally, we define the cross-mobility scale as
\begin{equation}
M_{\times}(\rho)
=
\frac{\chi_{\mathrm{mult}}}{k_B T}
\sqrt{D_r(\rho)D_\tau(\rho)}.
\label{eq:Mcross_main}
\end{equation}
This structure is central to the main result of this paper: even when the baseline positional
diffusivity is held fixed, varying the mobility in internal-state coordinates can
qualitatively reorganize motion in physical space through density-dependent transport and asymmetric cross-coupling.
\subsection{Cross-coupling and stochastic forcing}
\label{subsec:cross_coupling_main}

\noindent At the deterministic level, the cross-coupling enters through the density-curvature signal
$\widehat{\nabla^2\rho}_i$ and the amplitudes $M_{r\tau}$ and $M_{\tau r}$ appearing in
Eqs.~\ref{eq:r_tau_dot_main}. At the stochastic level, the increments $(\boldsymbol\eta_i,\zeta_i)$ are sampled from a local Gaussian covariance consistent with the mobility matrix. For one particle in two spatial dimensions, the covariance matrix has the generic form
\begin{equation}
\Sigma_i
=
2k_B T\,\Delta t
\begin{pmatrix}
M_r(\rho_i) & 0 & M_{\times}(\rho_i) \\
0 & M_r(\rho_i) & M_{\times}(\rho_i) \\
M_{\times}(\rho_i) & M_{\times}(\rho_i) & M_\tau(\rho_i)
\end{pmatrix},
\label{eq:noise_cov_main}
\end{equation}
subject to the positive-semidefinite condition
\begin{equation}
2\,M_{\times}(\rho_i)^2 \le M_r(\rho_i)M_\tau(\rho_i).
\label{eq:psd_constraint_main}
\end{equation}
The detailed sampling procedure, together with its relation to reciprocal passive limits and detailed
balance, is given in Appendix~\ref{app:noise_covariance}.
Because the off-diagonal terms in the deterministic flow are not equal, i.e., we do not have a symmetric gradient
flow, the present model is generically nonequilibrium even when the Gaussian noise is sampled
consistently with the local mobility matrix. The implications of this asymmetry, and its relation to
more reciprocal variants of the broader framework, are discussed in
Appendix~\ref{app:nonequilibrium_levels} and in the Discussion.

\subsection{Simulation protocol}
\label{subsec:simulation_protocol_main}

\noindent All simulations reported here are performed in a two-dimensional periodic square domain which is sufficiently large (for the relevant timescales we study) that minimum image convention for pair separation does not enter. Particles are initialized on a hexagonal lattice in spatial coordinates with $\tau_i(0)=0
 ~~\text{for all } i$.
We examine the role of different interaction regimes, $J_0$ and $J_2$, baseline mobility in internal coordinates $D_{\tau}^{0}$, dependence on local density $\beta_{\tau}$ and strength of cross-coupling through $\chi_{mult}$ and an additional scaling variable for the drift term, $g_{\chi}$. Finally, we also consider different system sizes, $N$, as indicative of the role of proliferation in such a context, though most of our results are for $N=1200$.
Unless otherwise stated, all simulations use the same interaction-law functional forms, the same intrinsic potential, the same kernel-based construction of local density and density curvature, and the same overdamped stochastic integration scheme. The fixed constants in this setup include $k_B T$, $\rho_c$, $a$, $r_c$, $w$, $\epsilon_{\mathrm{rep}}$, $\alpha_{\mathrm{rep}}$,  $\lambda$, the box size, and the time step, while the main control parameters varied across sweeps are $N$, $D_r^0$, $D_\tau^0$, $J_0$, $J_2$, $\beta_\tau$, $\chi$, $g_\chi$,
$\epsilon_{\mathrm{att}}$, and $\sigma_\tau$.
the fixed simulation constants are $a=1.0,L=80.0,r_c=3.0,
w=0.1$ for length scales, $k_B T=0.01,\epsilon_{\mathrm{rep}}=20.0,\epsilon_{\mathrm{att}}=0.2$ for energy scales, $\sigma_\tau=0.1,\lambda=10.0$ for the $\tau$-dependent potentials, density-dependence through $\rho_c=30,\beta_r=5$ and a timestep of $\Delta t=0.01$ with all simulations run for $n_{\mathrm{steps}}=10000$ steps. We keep the positional diffusivity, $D_r^0$ fixed at $D_r^0=0.01$.
Data are averaged over $4$ independent simulations unless otherwise specified.

\section{Results}
\label{sec:results}

\subsection{Interaction sectors set by $J_0$ and $J_2$}
\label{subsec:results_J0J2}

\begin{figure*}[t]
    \centering
    \includegraphics[width=\textwidth]{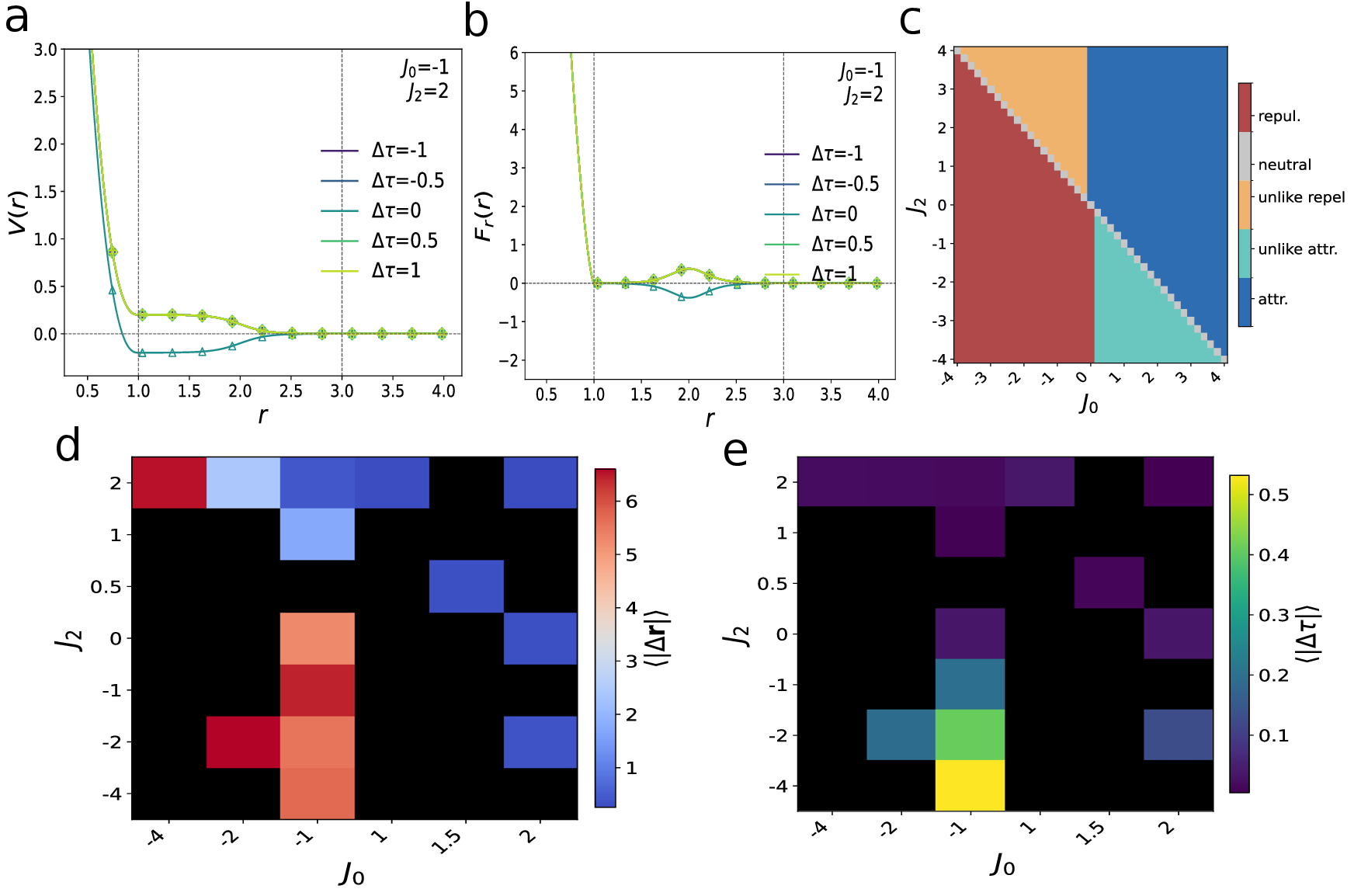}
    \caption{
    \textbf{Interaction sectors and their dynamical consequences.}
    (a) Pair potential $V(r)$ for a representative interaction choice
    $J_0=-1$, $J_2=2$, with $\epsilon_{\mathrm{att}}=0.2$ and $\sigma_\tau=0.1$, sweeping
    $\Delta\tau\in\{-1,-0.5,0,0.5,1\}$.
    (b) Corresponding radial force profile for the same parameter set.
    (c) Regime map in the $(J_0,J_2)$ plane at fixed $\epsilon_{\mathrm{att}}=0.2$ and
    $\sigma_\tau=0.1$, showing uniformly attractive, uniformly repulsive, and mixed sectors.
    (d,e) Summary heatmaps of long-time mean spatial and type displacements across representative
    interaction regimes at $N=1200$ for low baseline diffusivities, $D_r^0=0.01$, $D_{\tau}^0=0.001$ and $\beta_{\tau}=5$, averaged over $4$ independent simulations.
    }
    \label{fig:J0J2_overview}
\end{figure*}

We begin by isolating how the interaction parameters $J_0$ and $J_2$ structure the coupling between 
internal state and spatial coordinates. Figures~\ref{fig:J0J2_overview}a,b show that, for a representative
$(J_0,J_2)=(-1,2)$ choice, varying $\Delta\tau$ primarily modulates the shell attraction while
leaving the hard core unchanged. Figure~\ref{fig:J0J2_overview}c then maps the corresponding sectors
in the $(J_0,J_2)$ plane, separating uniformly attractive, uniformly repulsive, and mixed
regions where like-like leads to attraction or repulsion in low mobility dynamics, $D_r^0=0.01$, $D_{\tau}^0=0.001$ and $\beta_{\tau}=5$. These interaction sectors already translate into distinct dynamical regimes at the many-particle
level. In Figures.~\ref{fig:J0J2_overview}d,e, the strongest spatial excursions for low mobility occur in the negative $J_0$ sectors, indicating a different stable arrested configuration compared to the hexagonal initial arrangement, whereas the most compact behavior is obtained for the like-like-attractive branch.

\subsection{Type mobility controls the onset of spatial diffusion at fixed $D_r^0$}
\label{subsec:results_Dtau_beta}

\begin{figure*}[t]
    \centering
    \includegraphics[scale=0.5]{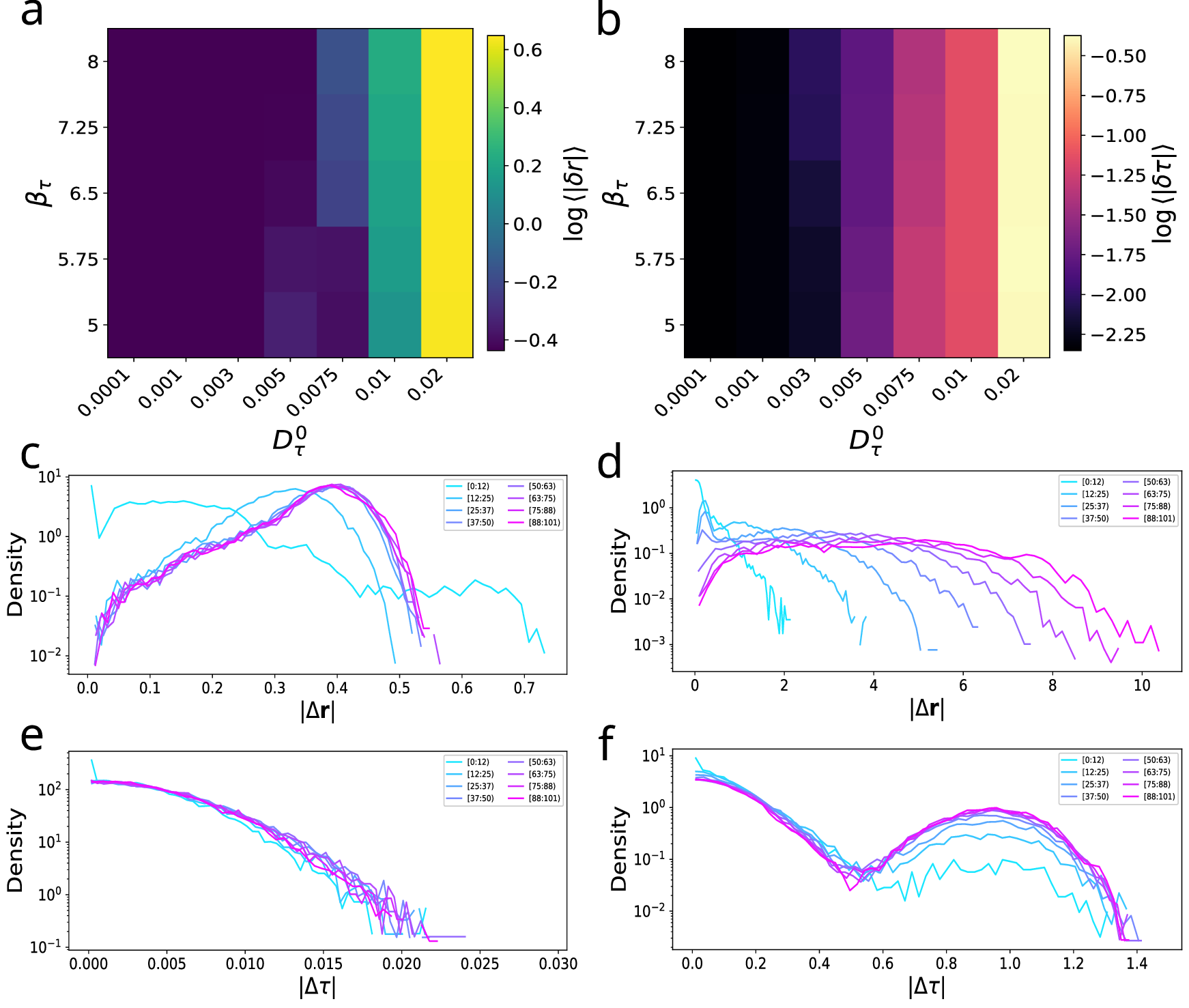}
    \caption{
    \textbf{Changing type mobility induces spatial diffusion at fixed $D_r^0$.}
    (a,b) Heatmaps of the mean spatial and type displacements over the
    $(D_\tau^0,\beta_\tau)$ sweep for the representative condition
    $D_r^0=0.01$, $J_0=-1$, $J_2=2$, $N=1200$, $\chi=0$.
    (c,d) Time-resolved distributions of $|\Delta r|$ for low and high $D_\tau^0$,
    respectively, at fixed $\beta_\tau=5$.
    (e,f) Corresponding distributions of $|\Delta\tau|$ for the same two cases.
    Low $D_\tau^0$ remains sharply localized, whereas high $D_\tau^0$ generates broad late-time
    tails in both sectors. 
    Data are averaged over $4$ independent simulations.}
    \label{fig:Dtau_beta_main}
\end{figure*}

We next turn to the main result of the paper: changing only the mobility in internal state can induce a qualitative reorganization of motion in real space even when the baseline positional
diffusivity $D_r^0$ is held fixed. Figure~\ref{fig:Dtau_beta_main}a shows that the late time mean spatial
displacement rises strongly as $D_\tau^0$ is increased, with a comparatively modest modulation
by $\beta_\tau$ when interactions are $J_0=-1,J_2=2$ (like-like attracts). This regime shows arrested behaviour at low mobility, with increasing mobility in internal coordinates inducing a change to migratory motion. Figure~\ref{fig:Dtau_beta_main}b shows the complementary trend in mean
absolute displacement in internal state, $\tau$. The temporal distributions make the transition especially clear. At low $D_\tau^0$,
Figures.~\ref{fig:Dtau_beta_main}c,e remain sharply concentrated near small $|\Delta r|$ and
$|\Delta\tau|$, with only weak broadening over time. By contrast, at high $D_\tau^0$,
Figures.~\ref{fig:Dtau_beta_main}d,f develop broad tails in spatial coordinates, indicating the onset of a late-time diffusive state, and the emergence of a second peak at $|\Delta\tau| = 1$, indicative of the transition from one well to another in $V_{intr}(\tau)$. The key point is that these large spatial excursions emerge without
changing the imposed baseline spatial diffusivity input.

\subsection{Co-movement in spatial and type coordinates}
\label{subsec:results_comovement}

\begin{figure}[t]
    \centering
    \includegraphics[width=\columnwidth]{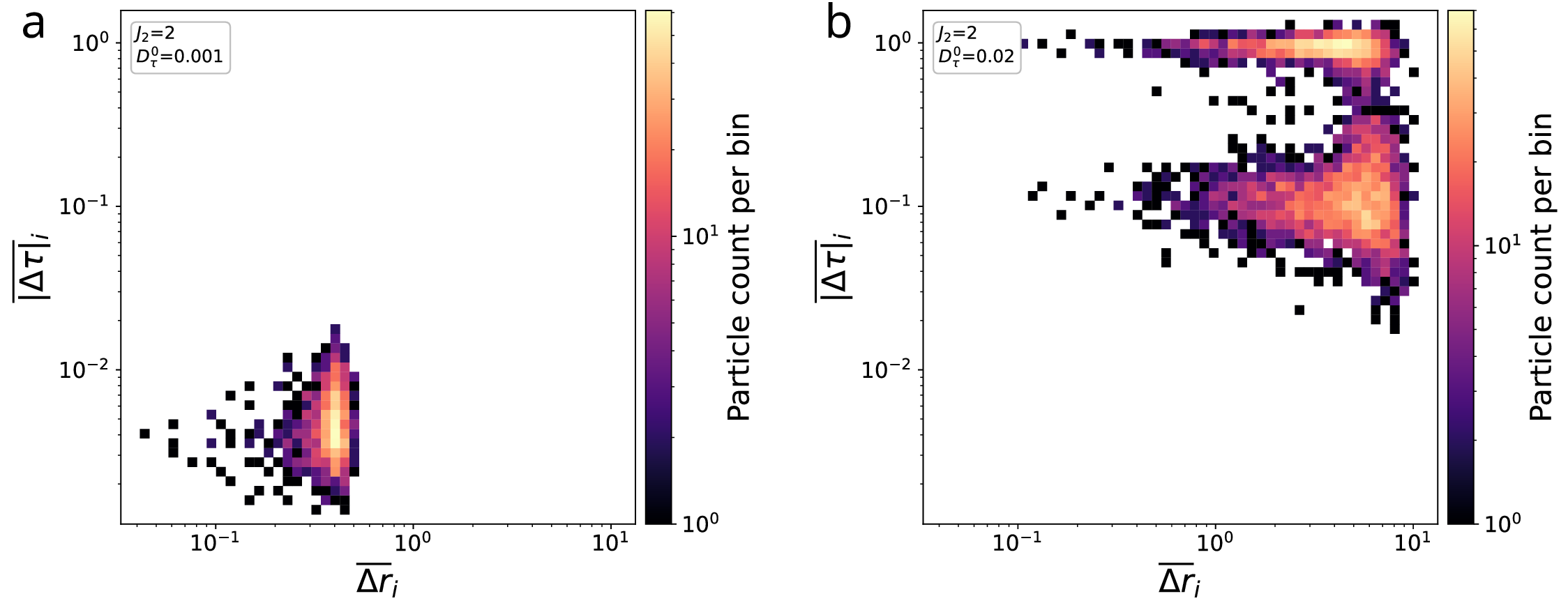}
    \caption{
    \textbf{Spatial and type displacements broaden together across the diffusive transition.}
    Joint histograms of spatial and type displacement for
    $D_r^0=0.01$, $J_0=-1$, $J_2=2$, $N=1200$, $\beta_\tau=5$, $\chi=0$.
    (a) Low-$D_\tau^0$ case, $D_\tau^0=0.001$, where the cloud remains tightly localized near the
    origin.
    (b) High-$D_\tau^0$ case, $D_\tau^0=0.02$, where both spatial and type excursions broaden
    strongly and show a clear positive co-movement.
    Data are averaged over $4$ independent simulations.}
    \label{fig:comovement_main}
\end{figure}

Having established that varying the internal state mobility reorganizes spatial motion, we next ask how
the two coordinates move jointly. Figure~\ref{fig:comovement_main} compares the joint distributions
of spatial displacement and type displacement for two representative $J_2=2$ conditions. At low
$D_\tau^0$ the particle cloud remains tightly concentrated near the origin
(Fig.~\ref{fig:comovement_main}a), indicating that both spatial motion and internal-state motion are
suppressed together. At high $D_\tau^0$, the cloud expands strongly and develops a clear positive bias (Fig.~\ref{fig:comovement_main}b), showing that larger spatial excursions are
typically accompanied by larger excursions in internal state, with analysis across $D_{\tau}^0$ suggesting a positive correlation (see Fig.~S5).
This result is important for the interpretation of the peeling-like regime below: the onset of
enhanced motion in real space is not independent of the internal-state sector, but is accompanied by
a correlated broadening in $\tau$.

\subsection{Density profiles and density curvature evolve with $D_\tau^0$ and time}
\label{subsec:results_density_curvature}

\begin{figure*}[t]
    \centering
    \includegraphics[width=\textwidth]{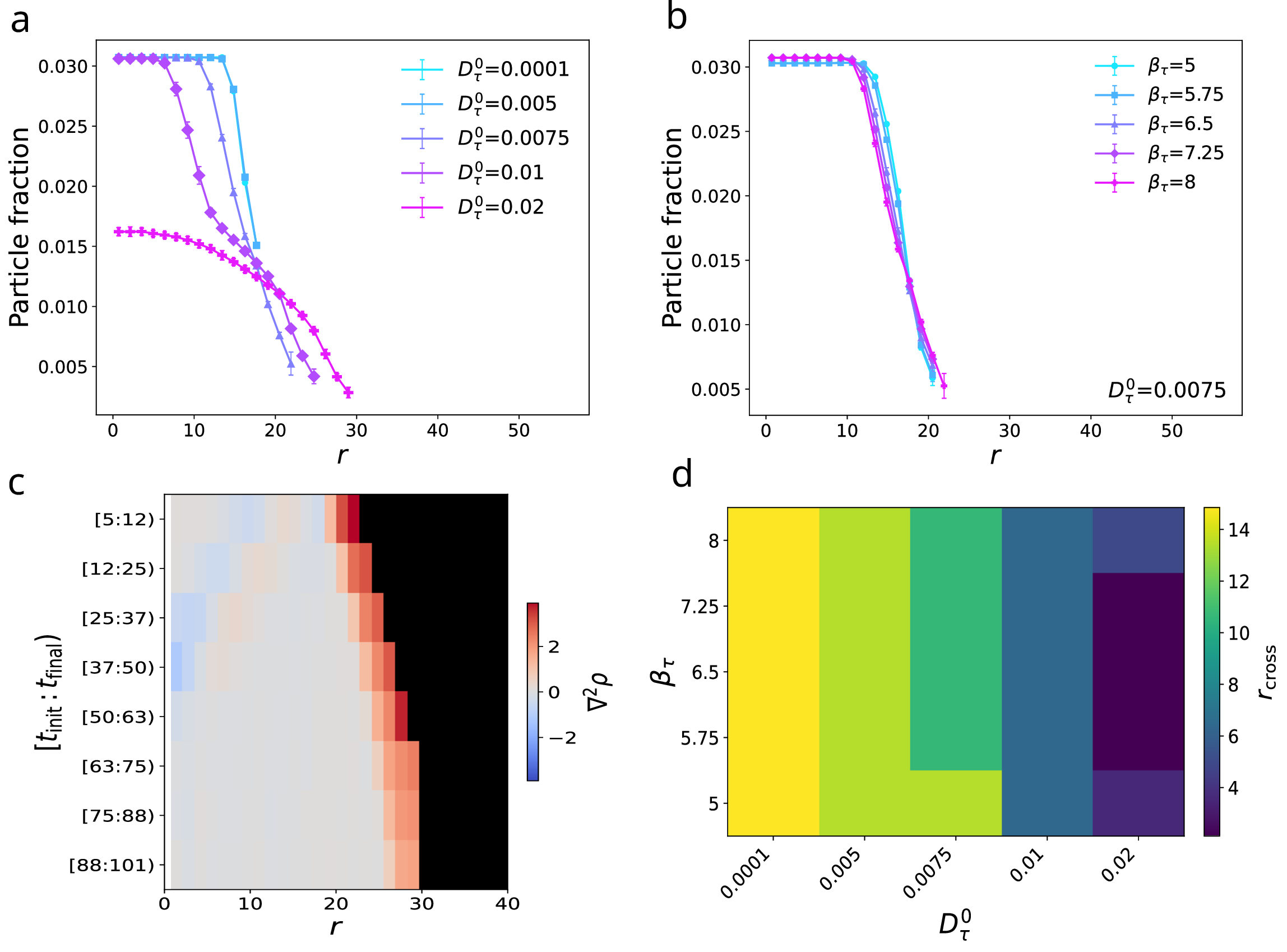}
    \caption{
    \textbf{Radial density profiles and density curvature reorganize across the transition.}
    (a) Radial particle-fraction profiles as $D_\tau^0$ is varied for the representative condition
    $D_r^0=0.01$, $J_0=-1$, $J_2=2$, $N=1200$, $\beta_\tau=8$, $\chi=0$.
    (b) Radial profiles over the $\beta_\tau$ sweep at fixed $D_\tau^0=0.0075$.
    (c) Representative time-resolved radial map of the density-curvature proxy.
    (d) Crossing radius,$r_{\mathrm{cross}}$, defined as the radial bin at which the density decreases by $10\%$ of the initial density at $r=0$, over the $(D_\tau^0,\beta_\tau)$ sweep, showing a
    strong collapse with increasing $D_\tau^0$ and weaker secondary dependence on $\beta_\tau$.
    Data are averaged over $4$ independent simulations.}
    \label{fig:density_curvature_main}
\end{figure*}

To connect the transition to local fields, we next examine the radial particle fraction and the
density-curvature proxy. Figure~\ref{fig:density_curvature_main}a shows that increasing
$D_\tau^0$ broadens the radial profile and shifts mass away from the most sharply concentrated core.
At an intermediate $D_\tau^0$, the dependence on $\beta_\tau$ remains present but is weaker than the
dominant dependence on $D_\tau^0$ itself (Fig.~\ref{fig:density_curvature_main}b).
The time-resolved density-curvature map in Fig.~\ref{fig:density_curvature_main}c shows the same
transition from a sharp, structured core--rim organization toward a flatter and more diffuse profile, with the region of large curvature change shifting over time.
This evolution is summarized in Fig.~\ref{fig:density_curvature_main}d through the crossing radius
$r_{\mathrm{cross}}$, measuring the radial bin at which the density decreases by $10\%$ of the initial density at $r=0$, which decreases markedly as $D_\tau^0$ increases. The dependence on
$\beta_\tau$ is comparatively weak at low $D_\tau^0$ and becomes visible mainly near the most
diffusive end of the sweep. Supplementary Figures. S1,S2
provide the full time-resolved curvature maps for representative
$\beta_\tau=5$ and $\beta_\tau=8$ cases.

\subsection{Representative snapshots of the coupled morphological and organizational states}
\label{subsec:results_snapshots}

\begin{figure*}[htpb!]
    \centering
    \includegraphics[width=\textwidth]{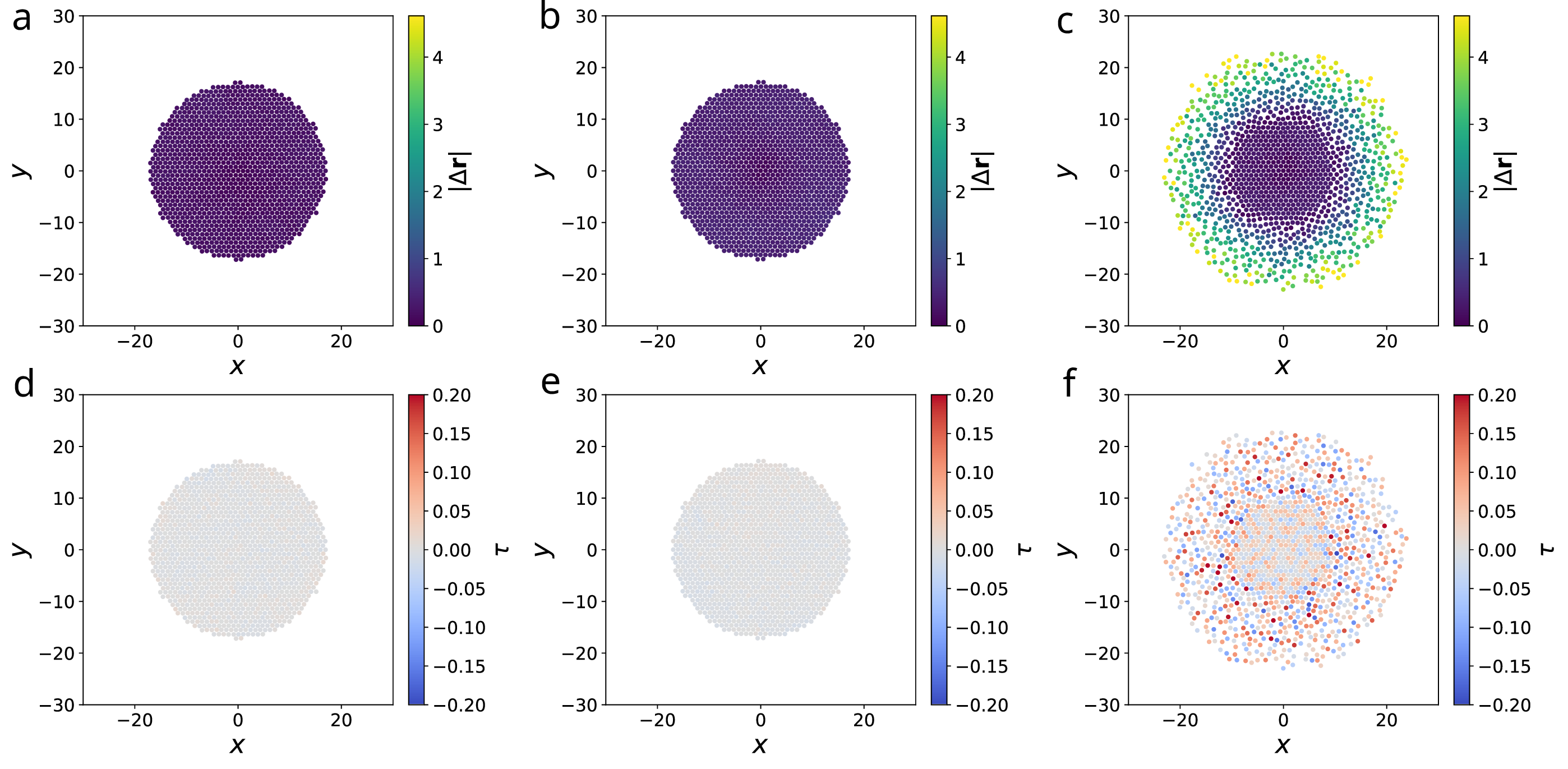}
    \caption{
    \textbf{Representative late-time morphologies I.}
    Top row: spatial maps colored by the magnitude of particle displacement $|\Delta r|$.
    Bottom row: corresponding internal-state field $\tau$.
    The panels show a representative progression from compact/arrested behavior to a pronounced
    peeling-like state in which the largest displacements are concentrated on the moving outer rim. The specific cases are $J_0=-1,J_2=2,D_r^0=0.02,D_{\tau}^0=0.001,\beta_{\tau}=8$ (a,d), $J_0=-1,J_2=2,D_r^0=0.01,D_{\tau}^0=0.001,\beta_{\tau}=5$ (b,e) and $J_0=-1,J_2=2,D_r^0=0.01,D_{\tau}^0=0.01,\beta_{\tau}=5$ (c,f). 
    }
    \label{fig:snapshots_set1}
\end{figure*}

\begin{figure*}[t]
    \centering
    \includegraphics[width=\textwidth]{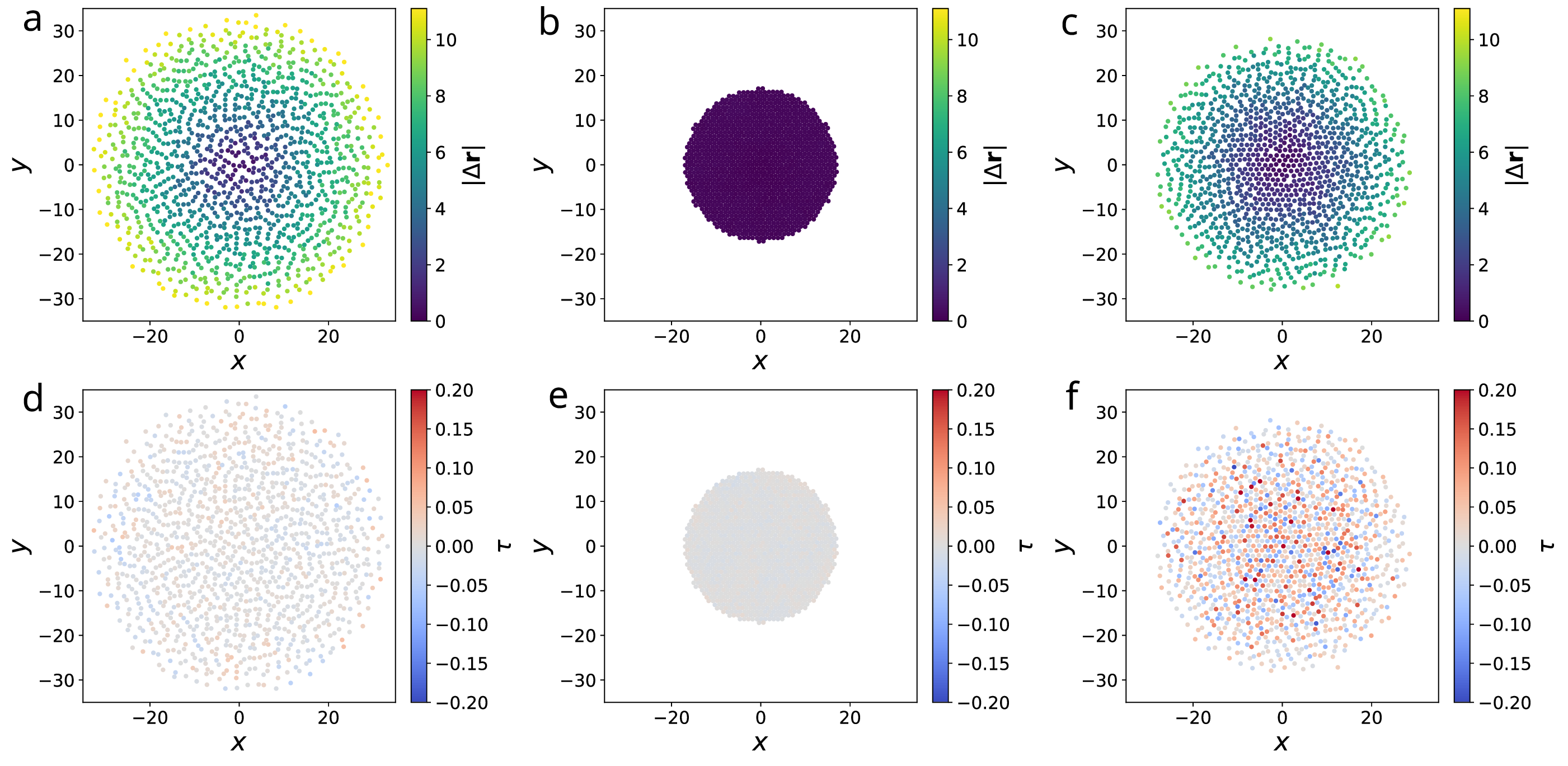}
    \caption{
    \textbf{Representative late-time morphologies II.}
    A second contrast set of representative final states from the same morphology sweep as in
    Fig.~\ref{fig:snapshots_set1}. Top row: spatial displacement magnitude $|\Delta r|$.
    Bottom row: internal-state field $\tau$.
    The comparison emphasizes that large spatial displacements remain concentrated at the advancing
    outer edge, while the dense core stays comparatively quiet in compact or arrested conditions.
    We show the following cases: $J_0=-4,J_2=2,D_r^0=0.02,D_{\tau}^0=0.001,\beta_{\tau}=5$ (a,d), $J_0=-1,J_2=2,D_r^0=0.02,D_{\tau}^0=0.001,\beta_{\tau}=8$ in (b,e) and $J_0=-1,J_2=1,D_r^0=0.01,D_{\tau}^0=0.02,\beta_{\tau}=5$ in (c,f)}
    \label{fig:snapshots_set2}
\end{figure*}

The preceding metrics indicate a transition from compact or arrested aggregates to states with much
larger late-time motion in both coordinate sectors for large internal state mobility. Figures~\ref{fig:snapshots_set1} and \ref{fig:snapshots_set2} make this
visually explicit. In both figures, the top row shows spatial maps colored by the magnitude of
particle displacement, while the bottom row shows the corresponding internal-state field.

In the compact and arrested cases, the interior remains nearly quiescent and the $\tau$ field stays
narrowly distributed. By contrast, in the peeling-like cases the largest spatial displacements are
localized on the outer rim, while the internal-state field broadens and becomes much more structured
near the boundary. This edge-localized activity is consistent with a migration-like peeling process
rather than uniform bulk fluidization. Supplementary Figures.~S3,S4
show the corresponding pair correlation maps, $g(r,\Delta\tau)$, for representative arrested and peeling cases, indicating internal-state-specific colocalisation for this regime of $J_0$, $J_2$.

In Fig.~\ref{fig:snapshots_set2} a,d, for $J_0=-4,~J_2=2$ which corresponds to a purely repulsive regime, one observes sparse structures reminiscent of spinodal decomposition, with a dense and less dense phase, and spatial heterogeneity in internal state.

\subsection{Motion in the diffusive regime is predominantly outward}
\label{subsec:results_outward_motion}

\begin{figure*}[t]
    \centering
    \includegraphics[scale=0.3]{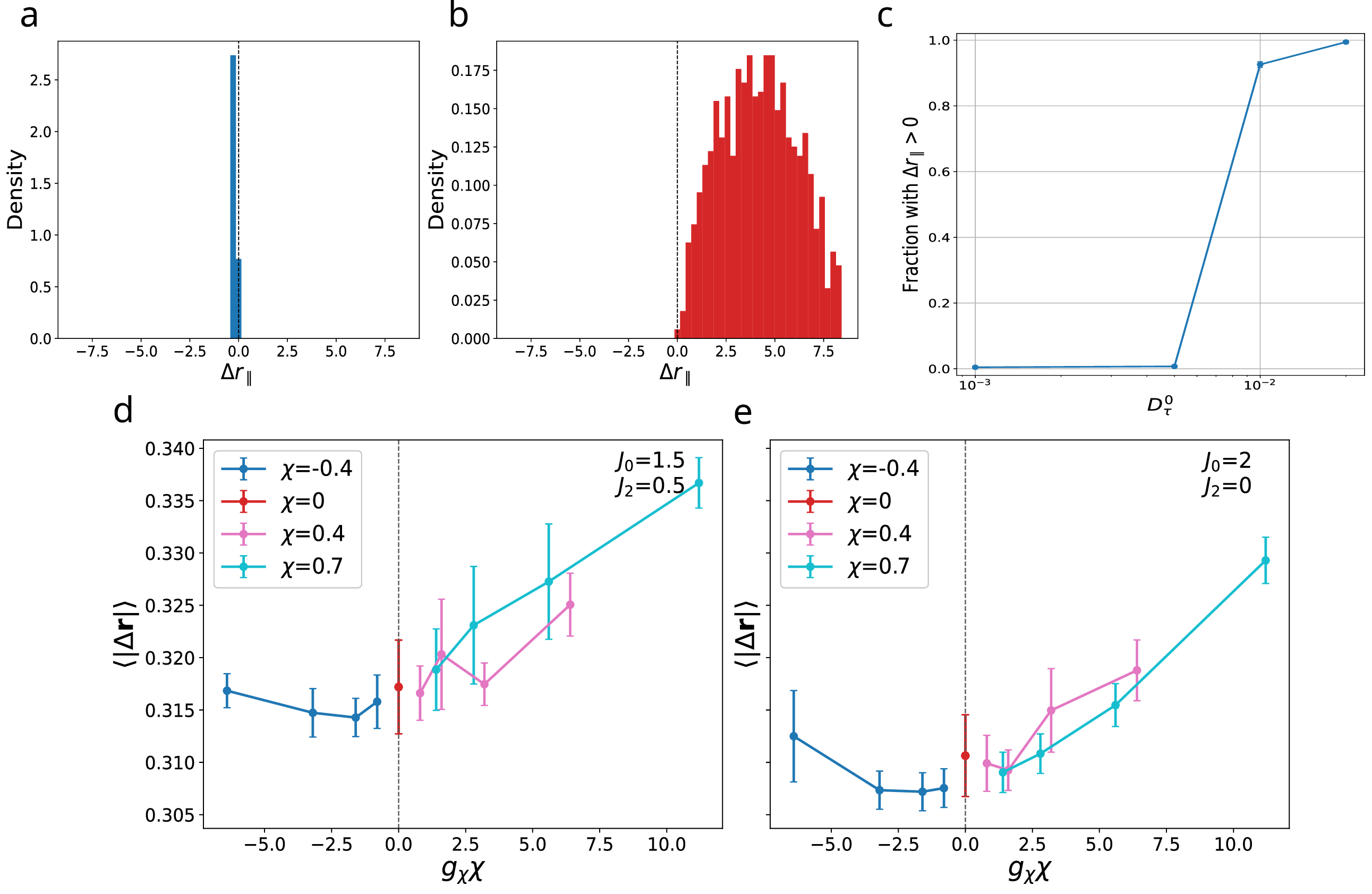}
    \caption{
    \textbf{The diffusive regime is dominated by outward motion.}
    (a,b) Histograms of signed radial displacement $\Delta r_i$ for representative arrested and
    peeling conditions from the outward-motion sweep with $J_0=-1$, $N=1200$, $\chi=0.7$,
$\epsilon_{\mathrm{att}}=0.2$, and $\sigma_\tau=0.1$. We use $D_{\tau}^{0}=0.001,\beta_{\tau}=8$ in (a) and $D_{\tau}^{0}=0.02,\beta_{\tau}=5$ in (b).
    (c) Fraction of particles with $\Delta r_i>0$ across the $D_\tau^0$ sweep for $\beta_{\tau}=8$, showing a sharp
    crossover from nearly zero to nearly unity.
    (d,e) Mean spatial displacement as a function of the effective coupling $g_\chi\chi$ for two
    representative interaction choices,
    $(J_0,J_2)=(1.5,0.5)$ and $(2,0)$, at
    $D_r^0=0.01$, $D_\tau^0=0.02$, $N=1200$, $\beta_\tau=4$,
    $\epsilon_{\mathrm{att}}=0.2$, and $\sigma_\tau=0.1$.
    Data are averaged over $4$ independent simulations.
    }
    \label{fig:outward_motion_main}
\end{figure*}

To quantify the directional character of the motion, we next examine signed radial displacements.
Figures~\ref{fig:outward_motion_main}a,b compare representative arrested and peeling cases. The
arrested case remains narrowly centered near zero with a slight inward or nearly neutral bias,
whereas the peeling case is strongly shifted to positive signed radial motion. Thus the diffusive
state identified above is not simply a high-variance state, but is dominated by outward migration
from the boundary.

Figure~\ref{fig:outward_motion_main}c summarizes this effect across the $D_\tau^0$ sweep through the
fraction of particles with $\Delta r_i>0$. The crossover is sharp and the outward-moving fraction stays
near zero at low $D_\tau^0$, then rises rapidly and approaches unity in the strongly diffusive
regime. Figure~\ref{fig:outward_motion_main} d,e show that once this regime has been entered, further changes in the effective
cross-coupling $g_\chi\chi$ produce only modest quantitative shifts in the mean spatial
displacement. The dominant control of the arrest-to-peeling transition therefore lies in the internal state mobility rather than in fine tuning of the effective cross-coupling.

\subsection{System size controls the onset and extent of peeling-like behavior}
\label{subsec:results_size_dependence}

\begin{figure*}[t]
    \centering
    \includegraphics[scale=0.4]{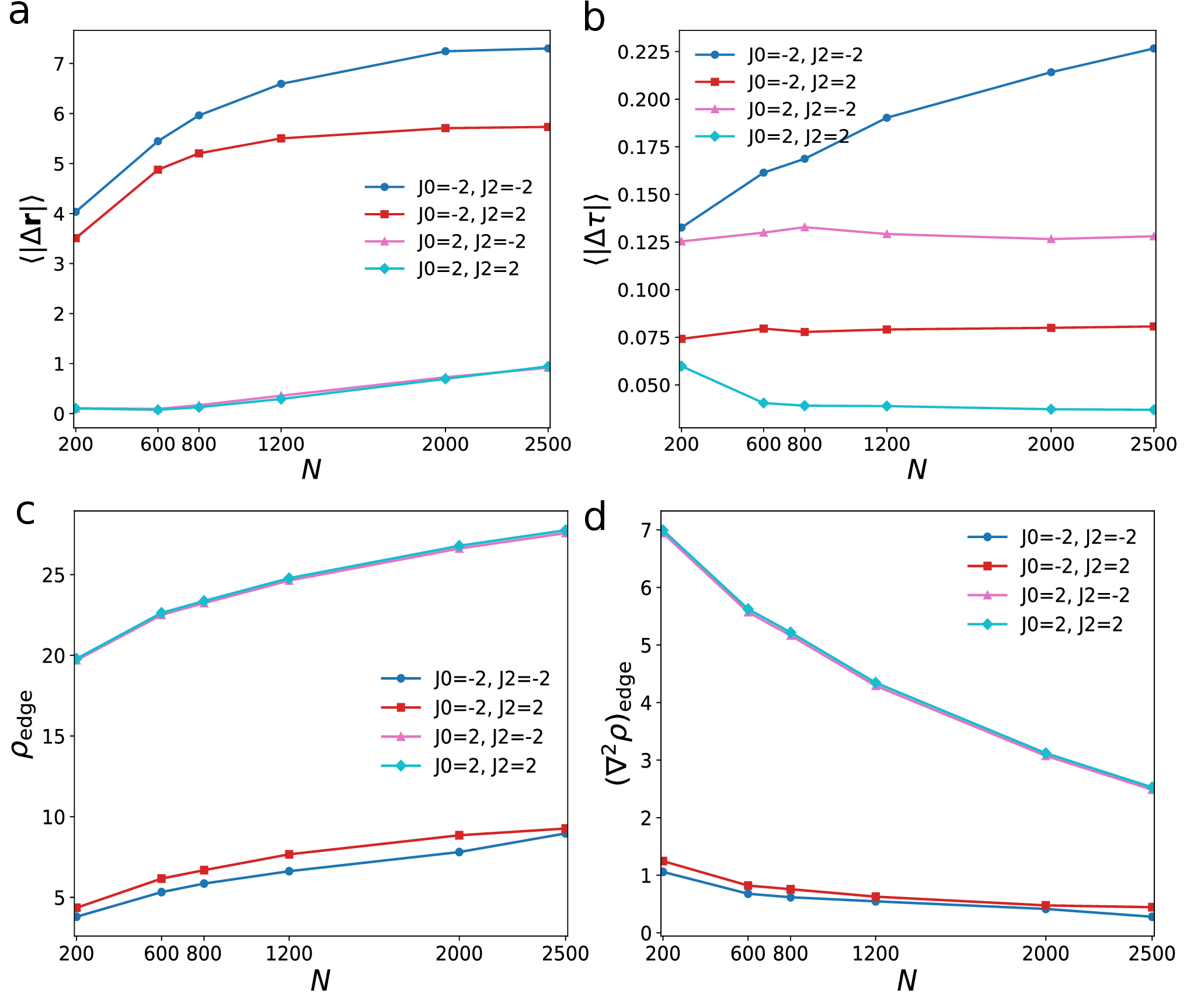}
    \caption{
    \textbf{System size modulates the onset of boundary-dominated migration.}
    (a,b) Mean spatial and type displacements across system size for representative interaction
    conditions from the size sweep with $D_r^0=0.01,~D_{\tau}^0=0.01,~\beta_{\tau}=8$ and
    $N\in\{200,600,800,1200,2000,2500\}$.
    The mobile negative-$J_0$ branches show a strong increase in spatial displacement with size,
    whereas the compact $J_0=2$ branches remain much less mobile.
    (c,d) Corresponding edge observables from the same size sweep, reporting the particle fraction at
    the edge and the edge density-curvature metric.
    Data are averaged over $4$ independent simulations.
    }
    \label{fig:size_dependence_main}
\end{figure*}

Finally, we examine how aggregate size modulates the onset of the peeling-like state. Figure
\ref{fig:size_dependence_main}a shows that the mean spatial displacement grows strongly with $N$ in
the mobile negative-$J_0$ branches, whereas the compact $J_0=2$ branches remain much less mobile
even at the largest sizes. Figure~\ref{fig:size_dependence_main}b shows the corresponding mean type
displacement, which remains comparatively small in the compact branches and grows more weakly than
the spatial sector. The edge observables in Figures.~\ref{fig:size_dependence_main}c,d show the same late-time ordering from a
boundary-focused perspective: the mobile branches develop a broader and more reorganized outer rim,
whereas the compact branches maintain a denser and less dynamically restructured edge. Together with
the line and time-course summaries in Supplementary Figures.~6,7
, this supports the interpretation that a growing aggregate can cross a
size threshold beyond which peeling-like, migration-dominated behavior becomes favorable even at
fixed microscopic interaction rules.
In particular, the $(J_0,J_2)=(-2,-2)$ and $(J_0,J_2)=(-2,2)$ conditions remain the most spatially mobile, while the
$J_0=2$ branches stay much more compact.
Chunked trajectories are shown in Supplementary Figures.~S6.

\section{Discussion}
\label{sec:discussion}

The central result of this work is that changing transport in internal coordinates can reorganize motion in
physical space without changing the baseline positional diffusivity input.
Primarily, this is driven by particles at the moving rim experience a different local spatial context than particles in the dense interior because the density-curvature signal is strongest  at the boundary. This creates a natural route to peeling-like outward motion even when the bare positional diffusivity $D_r^0$ is unchanged. 
Within this model, increasing $D_\tau^0$ and tuning $\beta_\tau$ drives a transition from
compact or arrested aggregates to a peeling-like regime with broadened spatial motion, strong
outward bias at the rim, and a marked size dependence. This identifies type-space mobility as an
independent control axis for collective reorganization, rather than a passive by-product of spatial
transport.
That conclusion is biologically suggestive. Many developmental and regenerative settings involve
coupled changes in state, neighbourhood, and tissue position rather than state progression in
isolation~\cite{oates2009quantitative,negrete2021towards,liu2024morphogenesis}. Likewise, migration
onset, epithelial reorganization, and patterning in stem-cell and organoid systems often involve
boundary-localized escape or rearrangement rather than homogeneous bulk motion
\cite{warmflash2014method,nousi2021single,alert2020physical,hallatschek2023proliferating}. Our results show that a minimal continuous state-space
augmentation is sufficient to generate this kind of qualitative transition and is therefore interesting as a modelling framework for such contexts.

Looking forward, such a description serves as a starting point to consider other specific contexts with well-defined additional ingredients. One direction is alignment and polarity~\cite{baconnier2025self}. Other promising directions include memory effects, where
history dependence can bias transitions and stabilize irreversible trajectories, and more explicit coupling of dynamics to geometry, curvature, and mechanics
\cite{hannezo2014growth,seferbekova2023spatial}. When coupled with systematic trajectory-based inference, such descriptions could have immense utility, and this first step of looking at the forward problem provides a basis to go further in the direction of connecting such principles to experimental data.

More broadly, we view this framework as a route toward mechanistic models that treat state change, spatial redistribution, and neighbourhood structure within a single framework of dynamics. We observe here that a minimal construction produces a nontrivial transition from arrest to peeling-like migration, suggesting that shared-coordinate modelling may provide a promising perspective and approach in the study of interacting populations and in biological systems undergoing development,
reprogramming, or disease-associated reorganization. 

\FloatBarrier



\eject
\part*{}   

\setcounter{equation}{0}
\setcounter{figure}{0}
\setcounter{section}{0}
\renewcommand{\theequation}{A\arabic{equation}}%
\renewcommand{\thefigure}{A\arabic{figure}}%
\renewcommand{\thesection}{Appendix A\arabic{section}}%


\appendix
\section*{Appendix}

\section{From the Hamiltonian to a continuity equation in shared coordinates}
\label{app:continuity_derivation}

\subsection{Empirical density and continuity equation}

Let the empirical density in the joint space of position and type be
\begin{equation}
\widehat\rho(\mathbf r,\tau,t)
=
\sum_{i=1}^N
\delta\!\bigl(\mathbf r-\mathbf r_i(t)\bigr)\,
\delta\!\bigl(\tau-\tau_i(t)\bigr).
\label{eq:rho_emp_app}
\end{equation}
Differentiating with respect to time gives
\begin{align}
\partial_t \widehat\rho
&=
-\sum_i
\dot{\mathbf r}_i\cdot\nabla_{\mathbf r}
\Bigl[
\delta\!\bigl(\mathbf r-\mathbf r_i\bigr)\,
\delta\!\bigl(\tau-\tau_i\bigr)
\Bigr]
-
\sum_i
\dot{\tau}_i\,\partial_\tau
\Bigl[
\delta\!\bigl(\mathbf r-\mathbf r_i\bigr)\,
\delta\!\bigl(\tau-\tau_i\bigr)
\Bigr]
\nonumber\\
&=
-\nabla_{\mathbf r}\!\cdot
\sum_i
\dot{\mathbf r}_i
\delta\!\bigl(\mathbf r-\mathbf r_i\bigr)
\delta\!\bigl(\tau-\tau_i\bigr)
-
\partial_\tau
\sum_i
\dot{\tau}_i
\delta\!\bigl(\mathbf r-\mathbf r_i\bigr)
\delta\!\bigl(\tau-\tau_i\bigr).
\label{eq:rho_emp_time_derivative_app}
\end{align}
Hence the empirical density obeys the exact conservation law
\begin{equation}
\partial_t \widehat\rho
=
-\nabla_{\mathbf r}\!\cdot \widehat{\mathbf J}_r
-\partial_\tau \widehat J_\tau,
\label{eq:continuity_emp_app}
\end{equation}
with empirical currents
\begin{equation}
\widehat{\mathbf J}_r(\mathbf r,\tau,t)
=
\sum_i
\dot{\mathbf r}_i\,
\delta\!\bigl(\mathbf r-\mathbf r_i\bigr)\,
\delta\!\bigl(\tau-\tau_i\bigr),
\qquad
\widehat J_\tau(\mathbf r,\tau,t)
=
\sum_i
\dot{\tau}_i\,
\delta\!\bigl(\mathbf r-\mathbf r_i\bigr)\,
\delta\!\bigl(\tau-\tau_i\bigr).
\label{eq:empirical_currents_app}
\end{equation}
Equation~\eqref{eq:continuity_emp_app} is purely kinematic and does not depend on the detailed form
of the dynamics.

\subsection{Free-energy functional associated with the microscopic Hamiltonian}

Starting from the broad Hamiltonian
\begin{equation}
\mathcal H
=
\sum_{i<j} U(\mathbf r_i-\mathbf r_j,\tau_i,\tau_j)
+
\sum_i V_{\mathrm{intr}}(\tau_i),
\label{eq:H_broad_app}
\end{equation}
the corresponding mean-field free-energy functional is
\begin{align}
\mathcal F[\rho]
&=
k_B T
\int d^2r\,d\tau\,
\rho(\mathbf r,\tau)
\bigl[\ln \rho(\mathbf r,\tau)-1\bigr]
\nonumber\\
&\qquad
+
\frac12
\int d^2r\,d\tau\,d^2r'\,d\tau'\,
\rho(\mathbf r,\tau)\,
U(\mathbf r-\mathbf r',\tau,\tau')\,
\rho(\mathbf r',\tau')
+
\int d^2r\,d\tau\,
\rho(\mathbf r,\tau)\,V_{\mathrm{intr}}(\tau).
\label{eq:F_nonlocal_app}
\end{align}
The first term is the ideal-mixing entropy, the second is the nonlocal interaction contribution, and
the third is the one-body potential for internal state. For the exact model used in this paper,
\begin{equation}
U(\mathbf r-\mathbf r',\tau,\tau')
\equiv
V\!\left(|\mathbf r-\mathbf r'|,\tau-\tau'\right),
\label{eq:U_exact_app}
\end{equation}
with $V$ given by Eqs.~\eqref{eq:V_pair_main}--\eqref{eq:C_offset_main} in the main text and
$V_{\mathrm{intr}}(\tau)$ given by Eq.~\eqref{eq:V_intr_main}. If desired, the nonlocal functional
can be expanded in gradients of $\rho$ to obtain a local Ginzburg--Landau form in the combined
space--type variables
by expanding the density about $(\mathbf r,\tau)$ in relative coordinates
\begin{equation}
\Delta \mathbf r = \mathbf r' - \mathbf r,
\qquad
\Delta \tau = \tau' - \tau.
\end{equation}
For a sufficiently smooth density field,
\begin{align}
\rho(\mathbf r',\tau')
&=
\rho(\mathbf r+\Delta\mathbf r,\tau+\Delta\tau)
\nonumber\\
&=
\rho
+
\Delta r_\alpha\,\partial_\alpha \rho
+
\Delta\tau\,\partial_\tau \rho
+
\frac{1}{2}\Delta r_\alpha \Delta r_\beta\,\partial_\alpha\partial_\beta \rho
+
\Delta r_\alpha \Delta\tau\,\partial_\alpha\partial_\tau \rho
+
\frac{1}{2}\Delta\tau^2\,\partial_\tau^2 \rho
+\cdots ,
\label{eq:rho_taylor_app}
\end{align}
where all derivatives on the right-hand side are evaluated at $(\mathbf r,\tau)$ and repeated
spatial indices are summed. Eq.~\ref{eq:rho_taylor_app} would be inserted into Eq.~\ref{eq:F_nonlocal_app}.

The associated chemical potential is
\begin{equation}
\mu(\mathbf r,\tau,t)
=
\frac{\delta \mathcal F}{\delta \rho(\mathbf r,\tau,t)}
=
k_B T \ln \rho(\mathbf r,\tau,t)
+
\int d^2r'\,d\tau'\,
U(\mathbf r-\mathbf r',\tau,\tau')\,
\rho(\mathbf r',\tau',t)
+
V_{\mathrm{intr}}(\tau).
\label{eq:chemical_potential_app}
\end{equation}

\subsection{Continuum currents and Onsager form}

At the continuum level, conservation in the shared coordinates takes the form
\begin{equation}
\partial_t \rho
=
-\nabla_{\mathbf r}\!\cdot \mathbf J_r
-
\partial_\tau J_\tau.
\label{eq:continuity_continuum_app}
\end{equation}
This corresponds to the following block mobility equation
\begin{equation}
\begin{pmatrix}
\mathbf J_r \\[3pt]
J_\tau
\end{pmatrix}
=
-\rho
\begin{pmatrix}
\mathbf M_{rr} & \mathbf M_{r\tau} \\[3pt]
\mathbf M_{\tau r} & M_{\tau\tau}
\end{pmatrix}
\begin{pmatrix}
\nabla_{\mathbf r}\mu \\[3pt]
\partial_\tau \mu
\end{pmatrix},
\label{eq:onsager_currents_app}
\end{equation}
where $\mathbf M_{rr}$ is a $2\times2$ mobility block, $M_{\tau\tau}$ is a scalar mobility in type
space, and the off-diagonal blocks couple the two sectors. In a reciprocal passive limit one expects
\begin{equation}
\mathbf M_{r\tau}
=
\mathbf M_{\tau r}^{\mathsf T}.
\label{eq:onsager_reciprocity_app}
\end{equation}
Eq.~\eqref{eq:onsager_currents_app} is the baseline general description of the dynamics.

\subsection{Connection to the microscopic equations used in this work}

The particle dynamics can be written as
\begin{equation}
\begin{pmatrix}
\dot{\mathbf r}_i \\[3pt]
\dot\tau_i
\end{pmatrix}
=
\begin{pmatrix}
\mathbf M_{rr}(\rho_i) & \mathbf M_{r\tau}(\rho_i) \\[3pt]
\mathbf M_{\tau r}(\rho_i) & M_{\tau\tau}(\rho_i)
\end{pmatrix}
\begin{pmatrix}
\mathbf F^{(r)}_i \\[3pt]
F^{(\tau)}_i
\end{pmatrix}
+
\boldsymbol{\xi}_i^{\mathrm{noise}},
\label{eq:micro_block_app}
\end{equation}
with
\begin{equation}
\mathbf F^{(r)}_i=-\frac{\partial \mathcal H}{\partial \mathbf r_i},
\qquad
F^{(\tau)}_i=-\frac{\partial \mathcal H}{\partial \tau_i}.
\label{eq:forces_micro_app}
\end{equation}

For the exact implementation used in the present manuscript, the conservative force sector is given
by Eqs.~\eqref{eq:F_r_main} and \eqref{eq:F_tau_main}, the local density field is estimated by
Eq.~\eqref{eq:rho_local_app}, and the cross-drift is driven by the density-curvature proxy
Eq.~\eqref{eq:laplacian_proxy_app}. In that sense the simulations should be viewed as a
microscopic, explicitly asymmetric realization of the broader shared-coordinate transport framework.


\section{Noise covariance, numerical sampling, and reciprocal passive limits}
\label{app:noise_covariance}

\subsection{Covariance matrix for one particle}

For one particle in two spatial dimensions, define the stochastic increment vector
\begin{equation}
\Delta\mathbf W_i
=
\begin{pmatrix}
\eta_{i,x} \\ \eta_{i,y} \\ \zeta_i
\end{pmatrix}.
\label{eq:noise_vector_app}
\end{equation}
In the most general, equilibrium case, the target covariance over one Euler--Maruyama step is
\begin{equation}
\left\langle
\Delta\mathbf W_i\,\Delta\mathbf W_i^{\mathsf T}
\right\rangle
=
\Sigma_i
=
2k_B T\,\Delta t
\begin{pmatrix}
M_r(\rho_i) & 0 & M_{\times}(\rho_i) \\
0 & M_r(\rho_i) & M_{\times}(\rho_i) \\
M_{\times}(\rho_i) & M_{\times}(\rho_i) & M_\tau(\rho_i)
\end{pmatrix}.
\label{eq:Sigma_full_app}
\end{equation}
Equivalently,
\begin{align}
\langle \eta_{i,\alpha}\eta_{j,\beta}\rangle
&=
2k_B T\,M_r(\rho_i)\,\Delta t\,
\delta_{\alpha\beta}\delta_{ij},
\\
\langle \zeta_i\zeta_j\rangle
&=
2k_B T\,M_\tau(\rho_i)\,\Delta t\,
\delta_{ij},
\\
\langle \eta_{i,\alpha}\zeta_j\rangle
&=
2k_B T\,M_{\times}(\rho_i)\,\Delta t\,
\delta_{ij},
\qquad \alpha\in\{x,y\}.
\end{align}
Setting the cross-covariances to zero and sampling the spatial and internal state increments independently introduces a non-equilibrium regime.

\subsection{Positive-semidefinite constraint and numerical draw}

The $3\times3$ covariance matrix in Eq.~\eqref{eq:Sigma_full_app} is positive semidefinite only if
\begin{equation}
2M_{\times}(\rho_i)^2 \le M_r(\rho_i)M_\tau(\rho_i).
\label{eq:psd_condition_app}
\end{equation}
In practice, the implementation enforces this by clamping $|M_{\times}|$ to a value just below the
bound,
\begin{equation}
|M_{\times}|
\le 0.999\,\sqrt{\frac{M_rM_\tau}{2}}.
\label{eq:clamp_app}
\end{equation}
Once the covariance matrix is constructed, the increment can be sampled by any standard Gaussian
factorization. Writing
\begin{equation}
\Sigma_i=\mathbf L_i\mathbf L_i^{\mathsf T},
\label{eq:chol_app}
\end{equation}
with $\mathbf L_i$ obtained from a Cholesky factorization (or an eigenvalue factorization when
needed for numerical robustness), one draws
\begin{equation}
\mathbf g_i\sim \mathcal N(\mathbf 0,\mathbf I_3),
\qquad
\Delta\mathbf W_i=\mathbf L_i\,\mathbf g_i.
\label{eq:noise_draw_app}
\end{equation}
The first two components are used as the spatial increment and the third as the type increment.

\subsection{Detailed balance: benchmark and caveat}

If the mobility matrix is reciprocal, the deterministic drift is derived from a common chemical
potential, and the noise covariance is matched to the same mobility through
Eq.~\eqref{eq:Sigma_full_app}, then one recovers the standard overdamped fluctuation--dissipation
structure. In the simplest constant-mobility case this yields an equilibrium measure of Gibbs form,
\begin{equation}
P_{\mathrm{eq}}(\{\mathbf r_i,\tau_i\})
\propto
\exp\!\left[-\frac{\mathcal H(\{\mathbf r_i,\tau_i\})}{k_B T}\right].
\label{eq:Gibbs_app}
\end{equation}
This reciprocal passive limit is the correct reference point for interpreting the model.

We make note of two caveats here. First, the production model uses
density-dependent mobilities while the noise terms have not been adjusted in any way to account for this, which would be required for exact detailed balance.
Second, the cross-terms in the drift components are coupled via the density-curvature-driven
active term in Eqs.~\eqref{eq:r_tau_dot_main}, and thus not by a symmetric force coupling. The model in this paper should therefore be understood as covariance-consistent at the level
of the sampled Gaussian increments, but not as an exact equilibrium sampler.


\section{Asymmetry and levels of nonequilibrium}
\label{app:nonequilibrium_levels}

\subsection{Reciprocal passive benchmark}

The first reference description is the reciprocal passive model. Here the currents in the shared
coordinates are generated by a single chemical potential $\mu$ and a reciprocal block mobility,
\begin{equation}
\mathbf M_{r\tau}
=
\mathbf M_{\tau r}^{\mathsf T},
\label{eq:reciprocal_app}
\end{equation}
with covariance-consistent noise. In this limit the spatial and type sectors are coupled, but the
coupling is still compatible with a generalized gradient flow of the free energy. This provides the
cleanest benchmark against which non-equilibrium descriptions can be compared.

\subsection{Density-dependent mobility without explicit active cross-drift}

A second layer of complexity arises when the mobilities depend on local density,
\begin{equation}
M_r=M_r(\rho),\qquad M_\tau=M_\tau(\rho),\qquad M_{\times}=M_{\times}(\rho).
\label{eq:density_dep_mob_app}
\end{equation}
Even before any explicit active asymmetry is introduced, this already creates a bulk--rim contrast in transport: dense regions are dynamically slowed relative to dilute regions. In the present
context this is important because a strong density dependence in internal state can make the interior
effectively arrested while leaving the outer rim comparatively mobile. This heterogeneous transport is
a necessary ingredient in the arrest-to-peeling transition highlighted in the main text.

\subsection{Asymmetric active cross-drift used in the present work}

The production model used for the figures departs more strongly from equilibrium. The deterministic
cross sector is not written as a matched force coupling between $\nabla_{\mathbf r}\mu$ and
$\partial_\tau\mu$, but rather through the density-curvature proxy
$\widehat{\nabla^2\rho}_i$:
\begin{align}
v_{\tau,i}^{\mathrm{cross}}
&=
g_\chi\,M_{r\tau}(\rho_i)\,k_B T\,\widehat{\nabla^2\rho}_i,
\\
\mathbf v_{r,i}^{\mathrm{cross}}
&=
g_\chi\,M_{\tau r}(\rho_i)\,k_B T\,
\widehat{\nabla^2\rho}_i\,
\widehat{\mathbf v}_{r,i}^{\mathrm{cons}}.
\end{align}
This breaks reciprocity in two distinct ways. First, the cross signal is a density-curvature field
rather than the gradient of a common chemical potential in the two sectors. Second, it enters through the structure of Eq.~\ref{eq:r_tau_dot_main} where the cross-drift enters the positional equation but not the type equation. The result is an intrinsically directional coupling from local density structure to spatial reorganization.

\section{Local density and density-curvature kernel construction}
\label{app:local_density_curvature}

To define density-dependent transport, we assign to each particle a local density and a local
density-curvature field obtained from a smooth finite-range neighborhood average. The basic kernel is
taken to be
\begin{equation}
\phi(r;r_c,w)
=
\frac12\left[1-\tanh\!\left(\frac{r-r_c}{w}\right)\right],
\label{eq:kernel_app}
\end{equation}
where $r_c$ sets the effective range of the neighborhood and $w$ controls the smoothness of the
cutoff. This kernel assigns large weight to nearby particles and suppresses contributions from
particles well beyond the cutoff radius.
The local density proxy at particle $i$ is then defined by
\begin{equation}
\rho_i
=
\phi(0;r_c,w)
+
\sum_{j\neq i}\phi(r_{ij};r_c,w),
\label{eq:rho_local_app}
\end{equation}
with $r_{ij}$ the minimum-image separation. In words, $\rho_i$ is a smooth count of nearby
neighbors, including a self-contribution at zero separation. Large $\rho_i$ therefore corresponds to
a locally crowded or bulk-like environment, while smaller $\rho_i$ indicates a more weakly populated
or boundary-like neighborhood.
To characterize whether a particle lies in the interior of a dense region or near a boundary, we
also construct a Laplacian-like density-curvature proxy. Let
\begin{equation}
M_0=\int \phi(r)\,d^2r,
\qquad
M_2=\int r^2\phi(r)\,d^2r,
\label{eq:kernel_moments_app}
\end{equation}
denote the zeroth and second isotropic moments of the kernel, evaluated numerically. The
density-curvature field used in the dynamics is then
\begin{equation}
\widehat{\nabla^2\rho}_i
=
\frac{4M_0}{M_2}
\frac{
\sum_{j\neq i}\phi(r_{ij};r_c,w)\,(\rho_j-\rho_i)
}{
\sum_{j\neq i}\phi(r_{ij};r_c,w)+\varepsilon
},
\label{eq:laplacian_proxy_app}
\end{equation}
where $\varepsilon$ is a small constant preventing division by very small weights.
This quantity measures the local curvature of the smoothed density field. Positive values correspond
roughly to particles lying in locally concave or bulk-centered regions of the density profile,
whereas negative values are associated with particles near outward-facing shoulders or boundaries.
For this reason, $\widehat{\nabla^2\rho}_i$ provides a natural scalar signal for distinguishing bulk
from rim environments. 
The same kernel construction can also be used to define other local fields, such as gradients of
smoothed internal-state averages.

\eject
\part*{}   

\setcounter{equation}{0}
\setcounter{figure}{0}
\setcounter{section}{0}
\renewcommand{\theequation}{S\arabic{equation}}%
\renewcommand{\thefigure}{S\arabic{figure}}%
\renewcommand{\thesection}{S\arabic{section}}%


\section*{Supplementary Figures}
\FloatBarrier

\begin{figure*}[h!t]
    \centering
    \includegraphics[width=\textwidth]{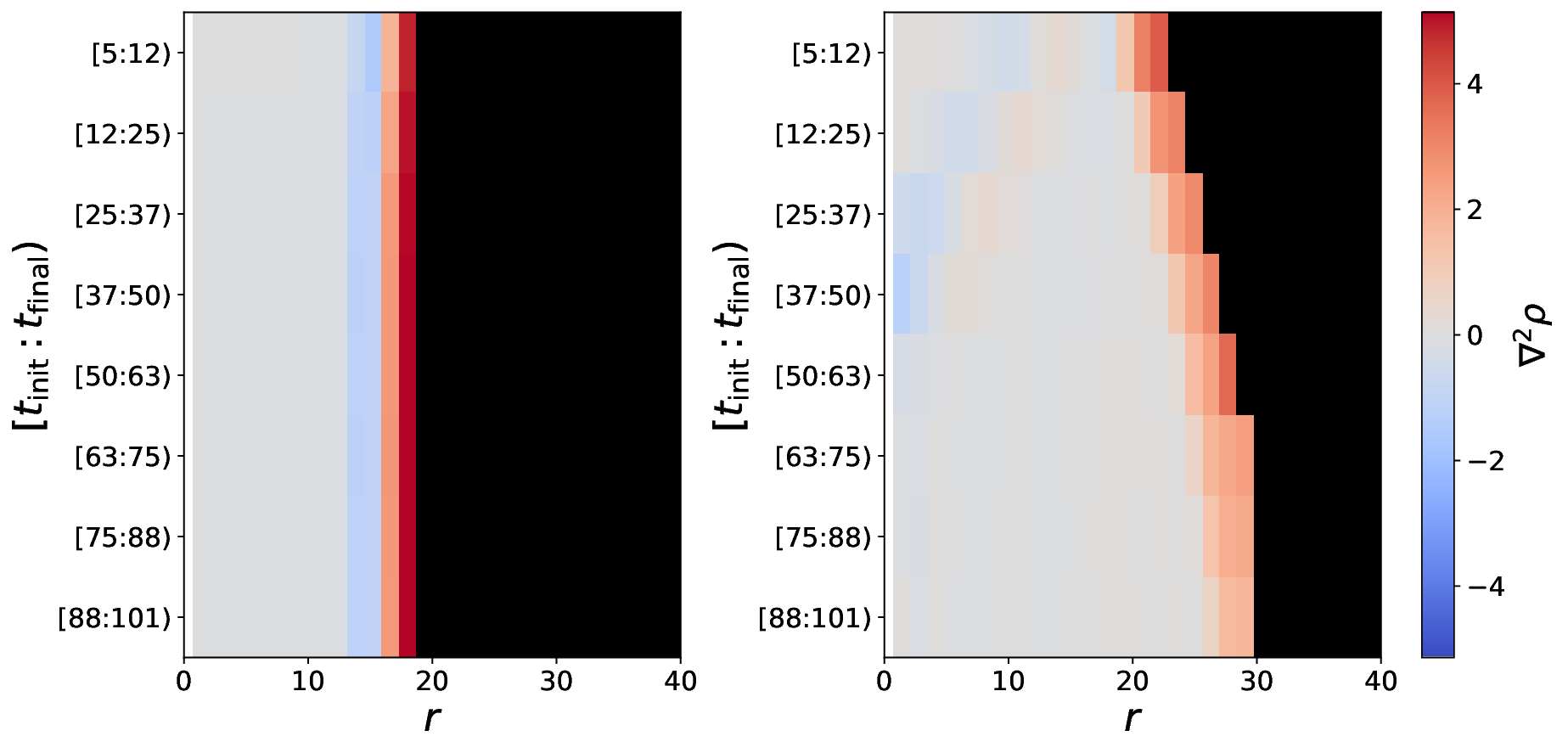}
    \caption{
    \textbf{Time-resolved density-curvature maps for $\beta_\tau=5$.}
    Radial heat maps of the density-curvature proxy for the representative condition
    $D_r^0=0.01$, $J_0=-1$, $J_2=2$, $N=1200$, $\beta_\tau=5$, $\chi=0$,
    $\epsilon_{\mathrm{att}}=0.2$, and $\sigma_\tau=0.1$, comparing
    $D_\tau^0=0.001$ (left) and $D_\tau^0=0.02$ (right).
    The low-$D_\tau^0$ case retains a sharper core--rim structure, whereas the high-$D_\tau^0$ case
    is visibly flatter and more diffuse at late times.}
    \label{figS:lap_time_beta5}
\end{figure*}

\begin{figure*}[t]
    \centering
    \includegraphics[width=\textwidth]{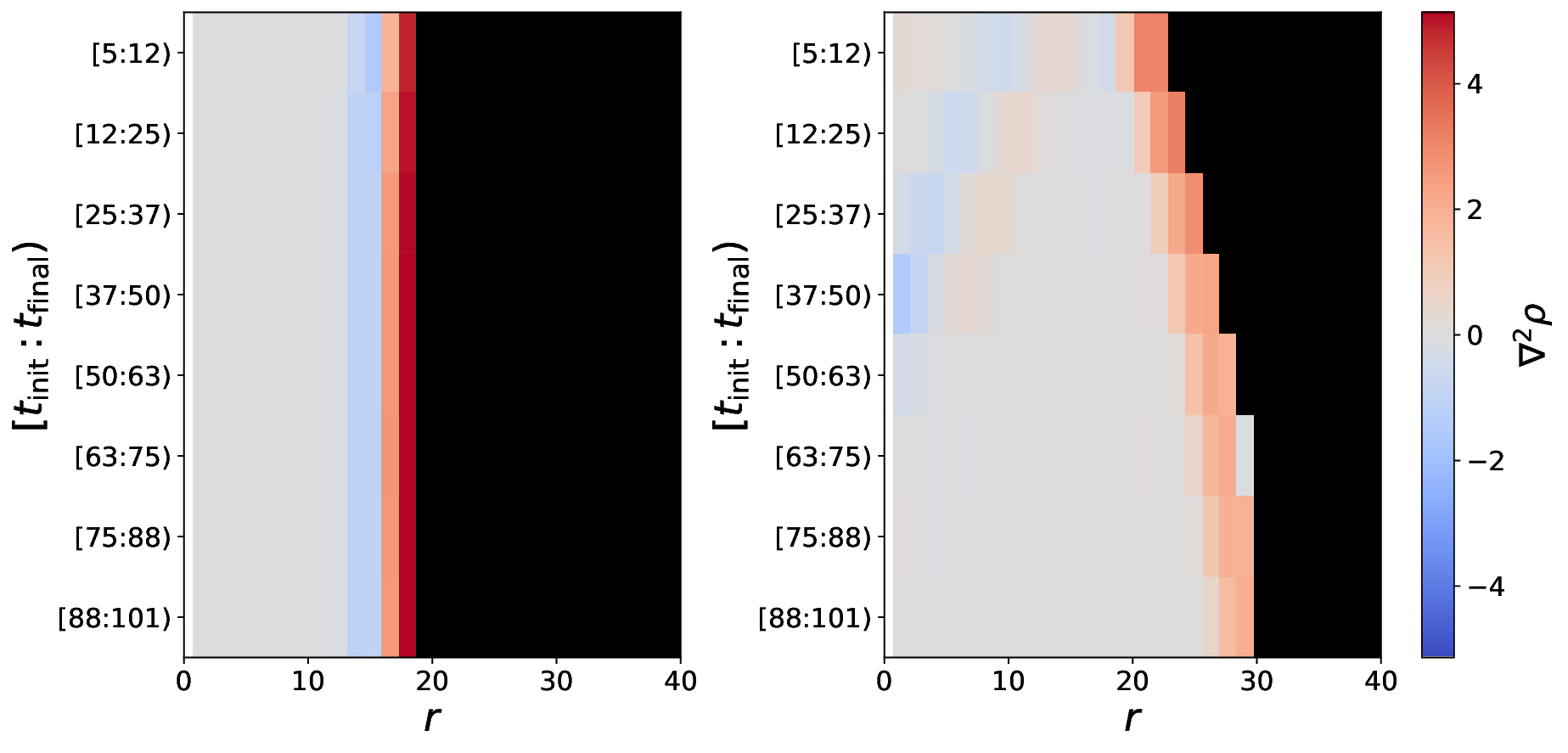}
    \caption{
    \textbf{Time-resolved density-curvature maps for $\beta_\tau=8$.}
    Same analysis as in Fig.~\ref{figS:lap_time_beta5}, with $D_{\tau}^0=0.001$ (left) and $D_{\tau}^0=0.02$ (right) but for $\beta_\tau=8$.
    The qualitative trend is unchanged: increasing $D_\tau^0$ smooths the density-curvature field and
    weakens the most extreme late-time structure, while the low-$D_\tau^0$ case preserves a sharper
    core--rim contrast.}
    \label{figS:lap_time_beta8}
\end{figure*}

\begin{figure*}[t]
    \centering
    \includegraphics[width=\textwidth]{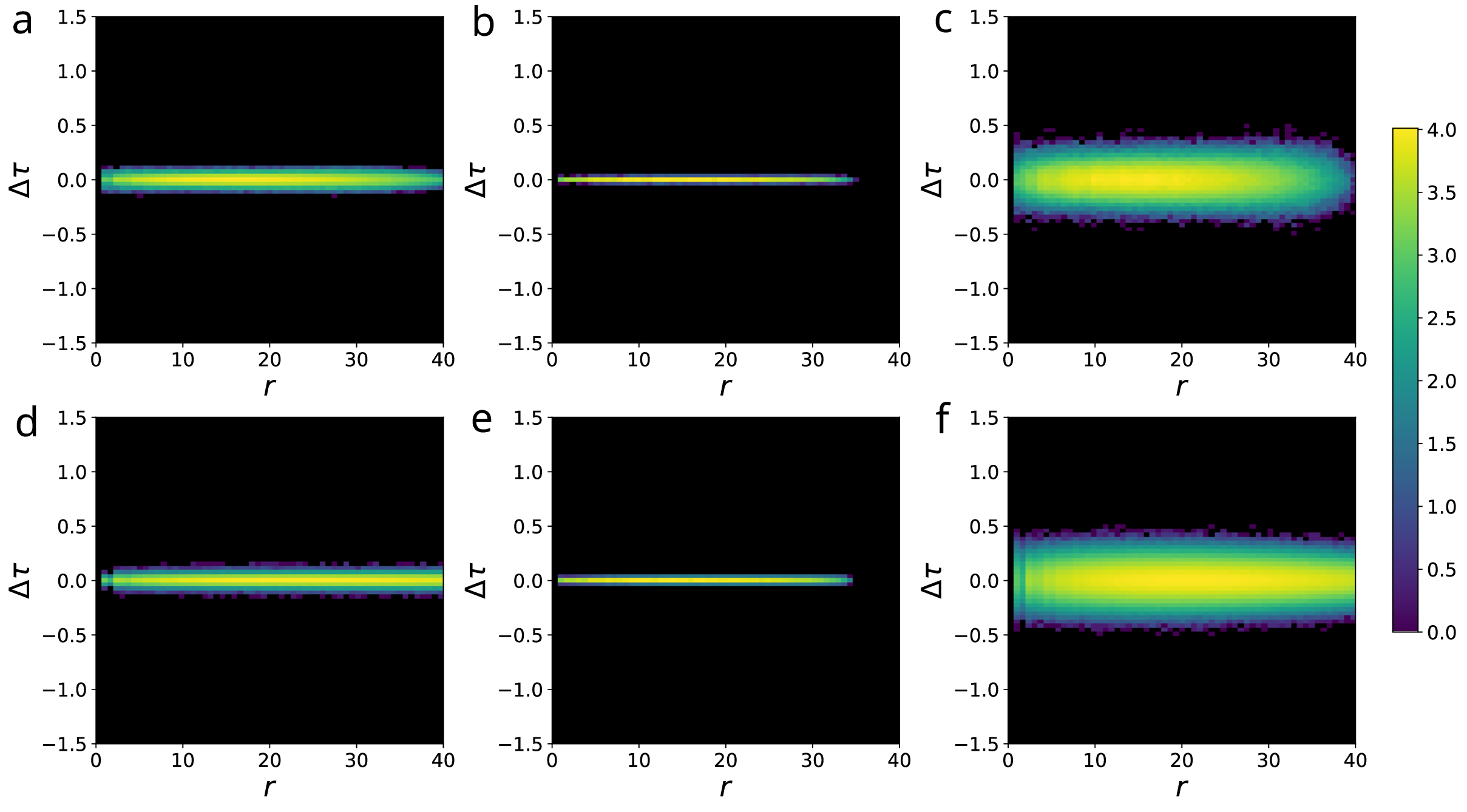}
    \caption{
    \textbf{Representative $g(r,\Delta\tau)_{\tau_{ref}}$ maps: contrast set I.}
    Joint radial--type distributions for representative conditions drawn from the morphology sweep at
    $N=1200$, with
    $D_r^0=0.01$,$D_\tau^0=0.001$, $J_0=-4$, $J_2=1$, $\beta_\tau = 5$, $\chi=0$ at initialisation (a) and in the final 500 time points (d), $D_r^0=0.01$,$D_\tau^0=0.001$, $J_0=-1$, $J_2=2$, $\beta_\tau = 5$, $\chi=0$ at initialisation (b) and in the final 500 time points (e),
    $D_r^0=0.01$,$D_\tau^0=0.02$, $J_0=-1$, $J_2=2$, $\beta_\tau = 5$, $\chi=0$ at initialisation (c) and in the final 500 time points (f).
    The arrested cases remain tightly localized near $\Delta\tau\simeq 0$, whereas the peeling-like
    cases broaden strongly in type space while retaining extended radial support.}
    \label{figS:gr_case_set1}
\end{figure*}

\begin{figure*}[t]
    \centering
    \includegraphics[width=\textwidth]{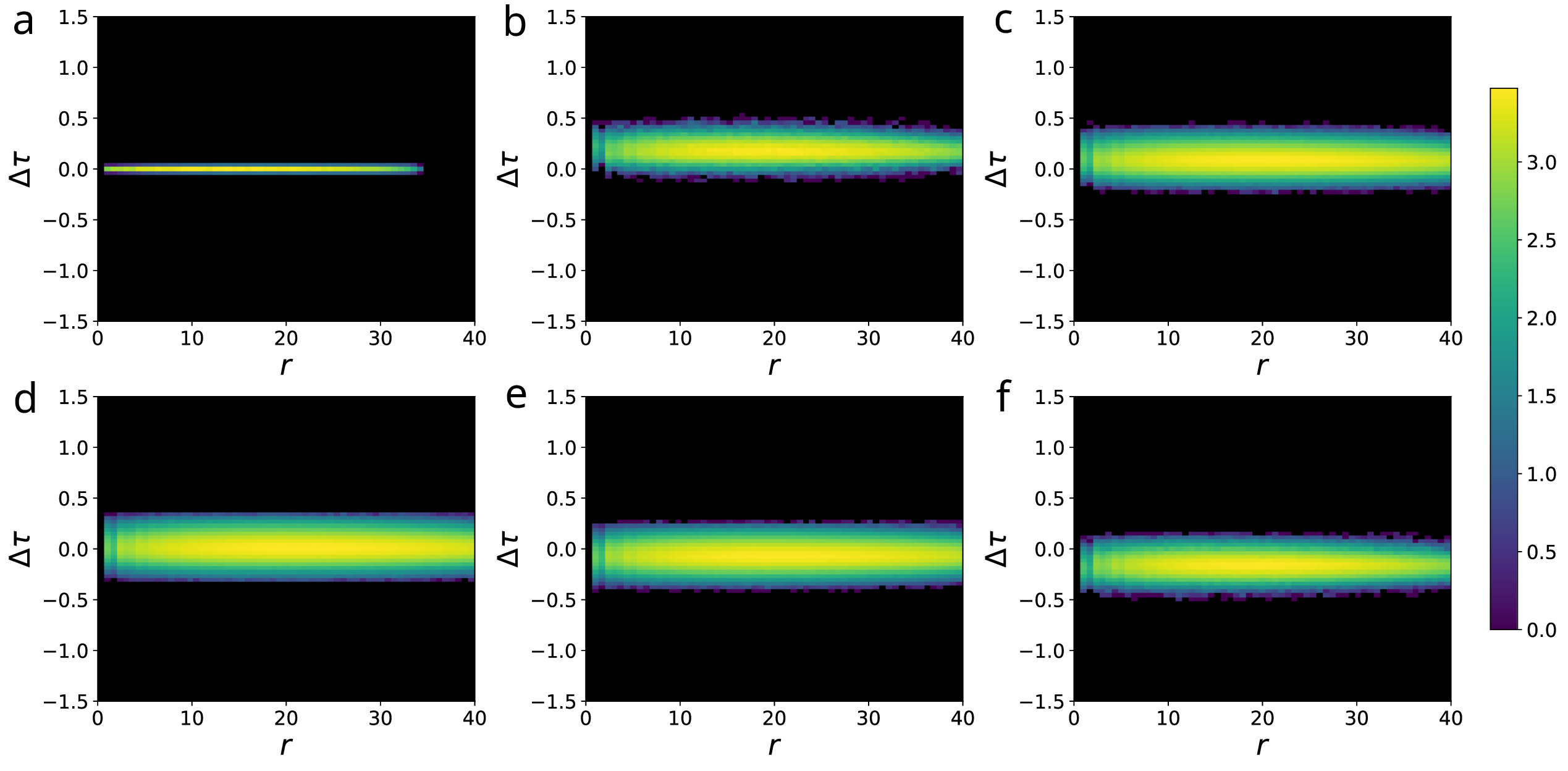}
    \caption{
    \textbf{Representative $g(r,\Delta\tau)_{\tau_{ref}}$ maps: contrast set II.}
    Additional joint radial--type distributions as in
    Fig.~\ref{figS:gr_case_set1} shown for different reference internal coordinate states, $\tau_{ref}$. The central reference particles are chosen such that their internal coordinate, $\tau$ is within $0.01$ of $\tau_{ref}$. The density map is then constructed with reference to this subset of particles.
    Here we show one arrest case (a) $J_0=-1, J_2=2, D_r^0=0.02, D_{\tau}^0=0.001, \beta_{\tau}=8, \chi=0.7$ at late timepoints for $\tau_{ref}=0$. We then show late time $g(r,\Delta \tau)_{\tau_{ref}}$ maps for $J_0=-1, J_2=1, D_r^0=0.01, D_{\tau}^0=0.02, \beta_{\tau}=5, \chi=0.7$ at $\tau_{ref}=-0.2$ (b), $-0.1$ (c), $0$ (d), $0.1$ (e), $0.2$ (f). }
    \label{figS:gr_case_set2}
\end{figure*}

\begin{figure*}[t]
    \centering
    \includegraphics[width=\textwidth]{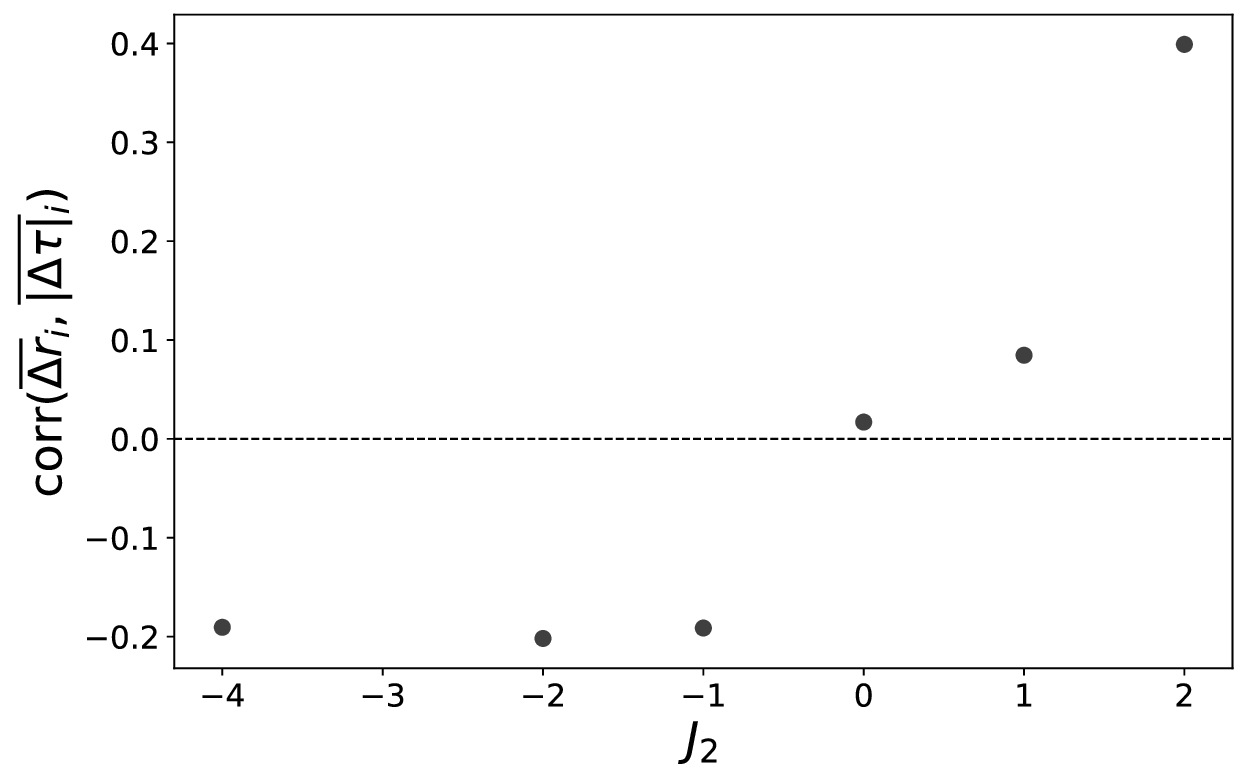}
    \caption{
    \textbf{Spatial--type co-movement changes sign across the $J_2$ sweep.}
    Replicate-level correlation between spatial displacement and type displacement for the
    representative condition
    $D_r^0=0.01$,
    $D_\tau^0=0.01$,
    $J_0=-1$,
    $N=1200$,
    $\beta_\tau=5$,
    $\chi=0$,
    and
    $J_2\in\{-4,-2,-1,0,1,2\}$.
    The correlation is moderately negative for $J_2=-4,-2,-1$, close to zero at $J_2=0$,
    positive at $J_2=1$, and strongest positive at $J_2=2$, showing that increasing positive
    $J_2$ promotes the clearest alignment between large spatial excursions and large type
    excursions.}
    \label{figS:comovement_v_J2}
\end{figure*}


\begin{figure*}[t]
    \centering
    \includegraphics[width=\textwidth]{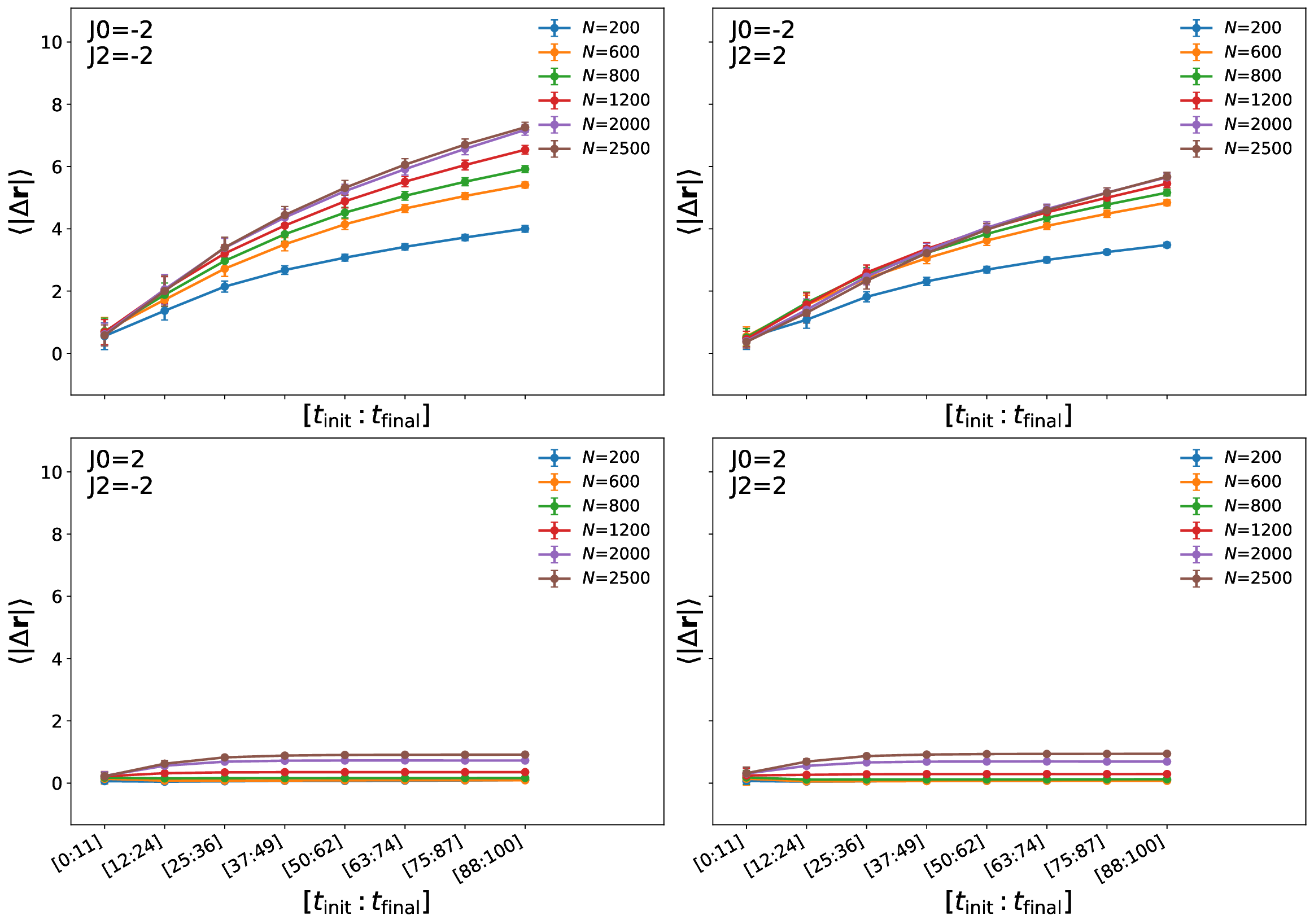}
    \caption{
    \textbf{Chunked trajectory growth across system size.}
    Time-resolved mean spatial displacement $\langle |\Delta r| \rangle$ for Parameters correspond to the size sweep with
    $D_r^0=0.01$,
    $D_{\tau}^{0}=0.001$,
    $(J_0,J_2)\in\{(-2,-2),(-2,2),(2,-2),(2,2)\}$,
    $\beta_\tau = 5$,
    $\chi = 0$. The data are resolved across chunked
    time windows and system sizes
    $N\in\{200,600,800,1200,2000,2500\}$.
    The negative-$J_0$ branches separate progressively over time, indicative of the overall repulsive regime, while the $J_0=2$ branches
    remain comparatively compact throughout the trajectory.}
    \label{figS:size_timecourses}
\end{figure*}

\end{document}